\documentclass{nature}
\usepackage[T1]{fontenc}
\usepackage[utf8]{inputenc}

\usepackage{graphicx}
\usepackage{epsfig}
\usepackage{hyperref}
\usepackage{threeparttable}
\usepackage{makecell}
\usepackage{xcolor}
\usepackage{ulem}
\usepackage{rotating}

\usepackage{graphicx}
\usepackage{epsfig}
\usepackage{amssymb}
\usepackage{amsmath}
\usepackage{mathrsfs}
\usepackage{color}
\usepackage[square,super,sort&compress]{natbib}

\usepackage[left]{lineno}
\usepackage{caption}
\usepackage[caption = false]{subfig}
\usepackage{amsmath, amssymb}

\newcommand{\lum}{erg~s\ensuremath{^{-1}}}

\newcommand{\ergs}{${\rm erg \ cm^{-2} \ s^{-1}}$ }

\newcommand\msun{\ifmmode M_{\odot} \else $M_{\odot}$\fi}
\newcommand\Lsun{\ifmmode L_{\odot} \else $L_{\odot}$\fi}
\newcommand{\lbol} {\ifmmode L_{\rm bol} \else $L_{\rm bol}$\fi}
\newcommand{\mbh}{\ifmmode M_{\rm BH} \else $M_{\rm BH}$\fi}
\newcommand{\ha}{\ifmmode {\rm H}\alpha \else H$\alpha$\fi}
\newcommand{\hb}{\ifmmode {\rm H}\beta \else H$\beta$\fi}
\newcommand{\vdisp}{\ensuremath{\sigma\mathrm{_{\star}}}}
\newcommand{\kms}{\ensuremath{\mathrm{km~s^{-1}}}}

\definecolor{orcidlogocol}{HTML}{A6CE39}

\def\be{\begin{eqnarray}}
\def\ee{\end{eqnarray}}

\makeatletter

\let\saved@includegraphics\includegraphics
\AtBeginDocument{\let\includegraphics\saved@includegraphics}
\renewenvironment*{figure}{\@float{figure}}{\end@float}
\makeatother

\makeatletter
\def\@fnsymbol#1{\ensuremath{\ifcase#1\or \dagger\or \ddagger\or
 \mathsection\or \mathparagraph\or \|\or **\or \dagger\dagger
 \or \ddagger\ddagger \else\@ctrerr\fi}}
\makeatother

\newcommand{\extfig}[1] {Extended Data Fig.}

\usepackage[utf8]{inputenc}
\usepackage{hyperref}
\usepackage{graphicx}
\usepackage{longtable}
\usepackage{supertabular}
\usepackage{float}
\begin{document}

\title{Tick-Tock: The Imminent Merger of a Supermassive Black Hole Binary }

\author{
Ning~Jiang$^{1,2}$\thanks{jnac@ustc.edu.cn}, Huan~Yang$^{3,4}$\thanks{hyang@perimeterinstitute.ca}, Tinggui~Wang$^{1,2}$, Jiazheng Zhu$^{1,2}$, Zhenwei Lyu$^{3,4}$, Liming Dou$^{5}$, Yibo Wang$^{1,2}$, Jianguo Wang$^{6}$, Zhen Pan$^{3}$, Hui Liu$^{1,2}$, Xinwen Shu$^{7}$, Zhenya Zheng$^{8}$
}

\maketitle
\begin{affiliations}
\item CAS Key Laboratory for Research in Galaxies and Cosmology, Department of Astronomy, University of Science and Technology of China, Hefei, Anhui 230026, China
\item School of Astronomy and Space Science, University of Science and Technology of China, Hefei 230026, China
\item Perimeter Institute for Theoretical Physics, Waterloo, ON N2L2Y5, Canada
\item University of Guelph, Department of Physics, Guelph, ON N1G2W1, Canada
\item Department of Astronomy, Guangzhou University, Guangzhou 510006, China
\item Yunnan Observatories, Chinese Academy of Sciences, Kunming 650011, China
\item Department of Physics, Anhui Normal University, Wuhu, Anhui, 241002, China
\item Shanghai Astronomical Observatory, 80 Nandan Road, Shanghai 200030, China
\end{affiliations}

\begin{abstract}

Supermassive black hole binaries (SMBHs) are a fascinating byproduct of galaxy mergers in the hierarchical universe\cite{Begelman1980}.  In the last stage of their orbital evolution, gravitational wave radiation drives the binary inspiral and produces the loudest siren\cite{Thorne1976,Haehnelt1994,Jaffe2003} awaiting to be detected by gravitational wave observatories. 
Periodically varying emission from active galactic nuclei has been proposed as a powerful approach to probe such systems\cite{Graham2015b,Liu2016,Charisi2016,Zheng2016,Chen2020}, although none of the identified candidates are close to their final coalescence such that the observed periods stay constant in time. 
In this work, we report on the first system with rapid decaying  periods revealed  by its optical and X-ray light curves, which has decreased from about one year to one month in three years. Together with its optical hydrogen line spectroscopy, we propose that the system is an  uneven mass-ratio, highly eccentric  SMBH binary which will merge within three years, as predicted by the trajectory evolution model. If the interpretation is true, coordinated, multi-band electromagnetic campaign should be planned for this first binary SMBH merger event observed in human history, together with possible neutrino measurements. Gravitational wave memory from this event may also be detectable by Pulsar Timing Array with additional five-to-ten year observation.

\end{abstract}


Close supermassive black hole binaries (SMBHBs) with separations below parsec (3.26 light years) scale are extremely challenging to find as they are beyond the resolution limit of current  generation telescopes except possibly with long baseline radio interferometry for very few nearby galaxies\cite{Gallimore2004}. Some indirect yet less conclusive methods have been proposed to search for subparsec SMBHBs, e.g. using the shift of broad emission lines by its orbit motion\cite{Eracleous2012,Shen2013,Runnoe2017} or the presence of double-peaked emission profiles from the coexisting broad-line regions associated with individual accreting SMBH\cite{Boroson2009}. The burgeoning time-domain surveys in the past decade have promoted another popular technique for unveiling SMBHBs, which is best demonstrated by the well-known  candidate OJ~287\cite{Sillanpaa1988,Lehto1996}. This methodology targets emissions from subparsec SMBHBs in active galactic nuclei (AGNs) that vary periodically on a time scale comparable to the orbit period, either driven by accretion rate fluctuations\cite{MacFadyen2008,Noble2012,Graham2015} or dominated by relativistic Doppler modulation\cite{D'Orazio2015}. Systematic searches for periodically varying AGNs have led to a considerable number of candidate SMBHBs with typical periods of years\cite{Graham2015b,Liu2016,Charisi2016,Zheng2016,Chen2020}. Assuming circular Keplerian orbits and BH masses of $10^8$~\msun, the separations of these candidate binaries correspond to centiparsecs to milliparsecs. In particular, the orbit of OJ287 is mainly driven by gravitational wave emission, but the expected merger is beyond Hubble time. 

\begin{figure}[h]
\centering
\includegraphics[width=16cm]{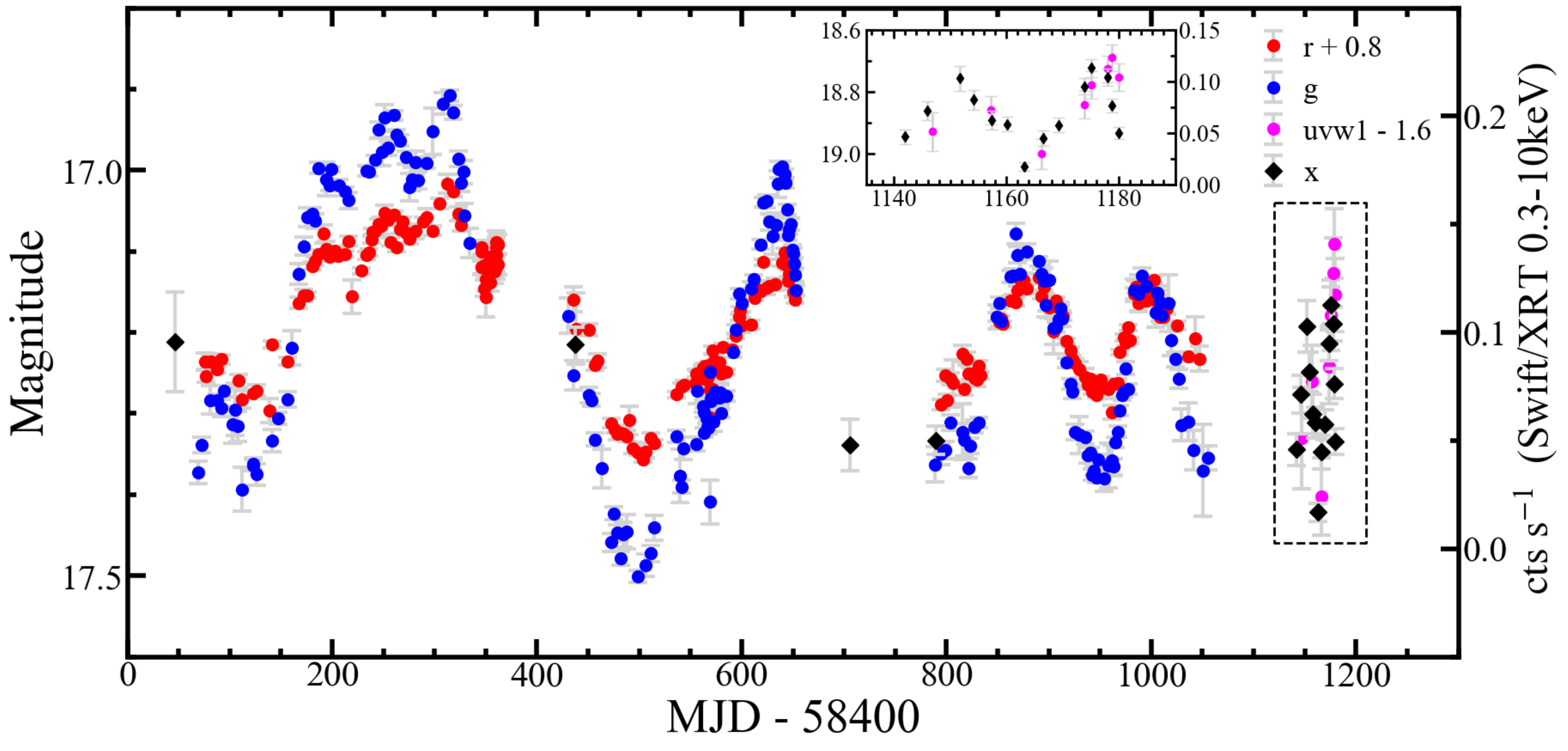}
\caption{
\textbf{The optical, UV and X-ray light curves of SDSSJ1430+2303. }
The ZTF $g$ and $r$ band photometric data are shown in blue and red solid circles, with error bars in grey. The black solid diamonds and magenta solid circles represent the XRT count rate in 0.3-10~keV and UVW1 magnitudes from our Swift monitoring, respectively.
We have zoomed in the Swift data (the region encircled by dashed box) in the inset for clarity.}
\label{lc}
\end{figure}

SDSSJ143016.05+230344.4 (hereafter SDSSJ1430+2303) is known as a Seyfert 1 galaxy at redshift 0.08105\cite{Oh2015}, which shows typical AGN-like narrow emission line ratios yet somewhat unusual blueshifted broad \ha\ emission (see Extended Data Fig.~\ref{sdsspec}) in its optical spectrum from Sloan Digital Sky Survey (SDSS). In particular, in the past three years the optical luminosity of SDSSJ1430+2303 has shown an unprecedented time-dependent variation. The $g$ and $r$ band light curves from Zwicky Transient Facitity (ZTF)\cite{Bellm2019} display an oscillation pattern since early 2019, which has completed at least 3 cycles up to August 2021,  with a decreasing oscillation amplitude and period (see Figure~\ref{lc}). To our best knowledge, such a chirping AGN with simultaneously  rapid  damping amplitude and period has never been reported in the past. Unfortunately, the target is invisible in the subsequent months because it is too close to the solar direction.  We have triggered the Neil Gehrels Swift X-ray telescope (XRT)\cite{Burrows2005} monitoring on this intriguing target since Nov. 24, 2021 (see details of Swift observations in Methods) and discovered a further shortened periodic variation in X-ray bands, whose period is approximately one month at the end of 2021.

The chirping flares are not compatible with known disk oscillation/instabilities, which have been tentatively used to explain other recurring AGN variabilities such as quasi-periodic eruptions\cite{Miniutti:2019fqr} and quasi-periodic oscillations \cite{gierlinski2008periodicity}.
The rapid decaying periods, which has decreased from $\sim1$ year to $\sim1$ month within only three years, also disfavour dissipation mechanisms such as dynamical friction at accretion disk crossings and/or tidal gravitational heating of stars near pericenter passages as the main drivers for orbital evolution (see details in alternative model part in Methods). It appears the only plausible scenario is a secondary black hole orbits around the primary SMBH in an inclined, highly eccentric trajectory. The secondary black hole crosses the accretion disk shortly before and after the pericenter passages, where significant energy and angular momentum are radiated away through gravitational waves (see trajectory model part in Methods), and the induced shock waves at disk-crossings eject plasma balls to produce observed flares in the optical band. Note that the flare luminosity is on the same order of magnitude as the background disk luminosity, indicating that the mass ratio between these two black holes cannot be too extreme. The X-ray emission from hot corona around the SMBH(s) are  likely affected by the pericenter passages, where the direct accretion onto black holes are mostly perturbed, but they may be also subject to variations in other circum-single  disk conditions.

\begin{figure}[h]
\centering
\includegraphics[width=10cm]{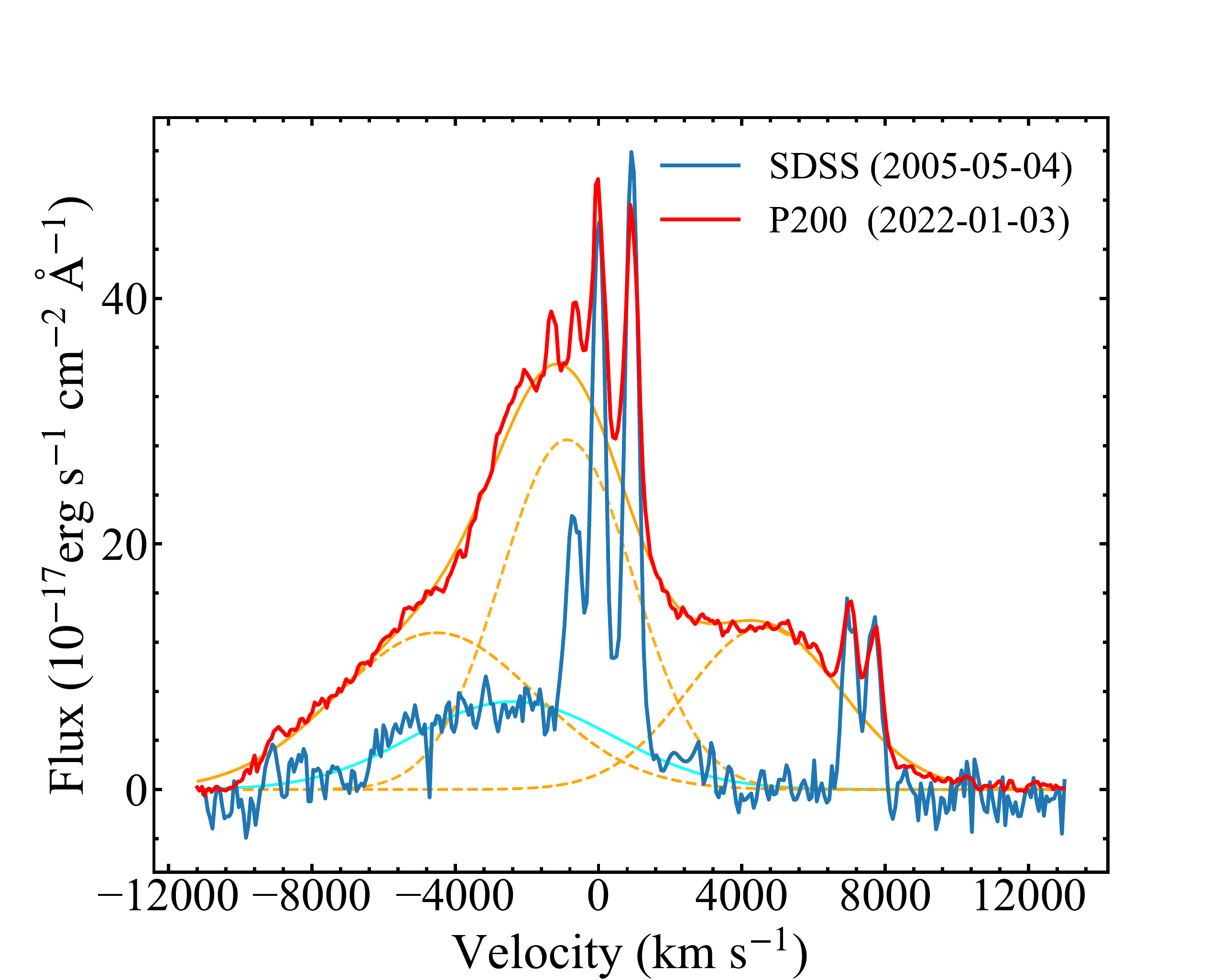}
\caption{
\textbf{The spectroscopic profile change around \ha\ region.} The broad \ha\ in SDSS spectra (blue, taken in 2005) is blueshifted with velocity of $\sim2400$~\kms\ (cyan line). The very recent spectra from P200/DBSP obtained on 2022 Jan. 3 (red) has shown a complex velocity structure, which can be fitted with three Gaussians (orange dashed lines), including a significantly redshifted ($\sim4600$~\kms) and a blueshifted ($\sim4000$~\kms) component.
}
\label{ha}
\end{figure}

The SMBHB scenario is further supported by the SDSS spectrum obtained more than one decade ago. First, the broad \ha\ emission line is oddly blueshifted  with a velocity of 2400~\kms\ relative to the whole galaxy, namely the reference system defined by narrow lines (see Figure~\ref{ha}). On the other hand, the SMBH mass (\mbh) estimated from broad \ha\ is $\sim4\times10^7$ times of solar mass (\msun) using the empirical virial mass estimator for single-epoch AGN spectra\cite{Greene2005}. However, the predicted \mbh\ is significantly smaller than that given by the tight scaling relations between SMBHs and their host galaxy properties\cite{Kormendy2013}. For instance, the stellar velocity dispersion predicts a mass of $\sim2\times10^{8}$~\msun, which will be even higher  derived from the total stellar mass. The mass discrepancy is no longer an issue  in the SMBHB scenario, e.g. the less massive secondary SMBH was actively accreting while the more massive primary SMBH was quiescent when SDSS spectra was taken. The secondary SMBH, carrying its broad-line region (BLR) clouds, was orbiting around the primary SMBH, with a projected  velocity of 2400~\kms\ as suggested by the Doppler shift of broad \ha\ line. 
If that is the case, the current separation between the SMBH binary has reduced to a distance much smaller than the original BLR size ($\sim7$ light days, see optical spectra part in Methods), resulting in a circum-binary BLR. During the approaching process,  the BLR clouds might be tidally disturbed and scattered by the primary SMBH, resulting in additional high-velocity unbound components. This scenario is well consistent with the distinct Balmer emission-line profiles revealed by recent optical spectroscopic observations - they all show multiple velocity structures, i.e. both obviously blueshifted  and redshifted components with velocity of thousands of~\kms\ (see Figure~\ref{ha} and more details in optical spectra follow-up part in Methods). In addition,  the abrupt outburst occurred between 2014 and 2017 (see Extended Data Fig.~\ref{oxlc}) may be explained as the accretion gas originally bound in the secondary black hole being captured by the primary black hole and feeding its accretion flow.
 
 \begin{figure}[h]
\centering
\includegraphics[width=16cm]{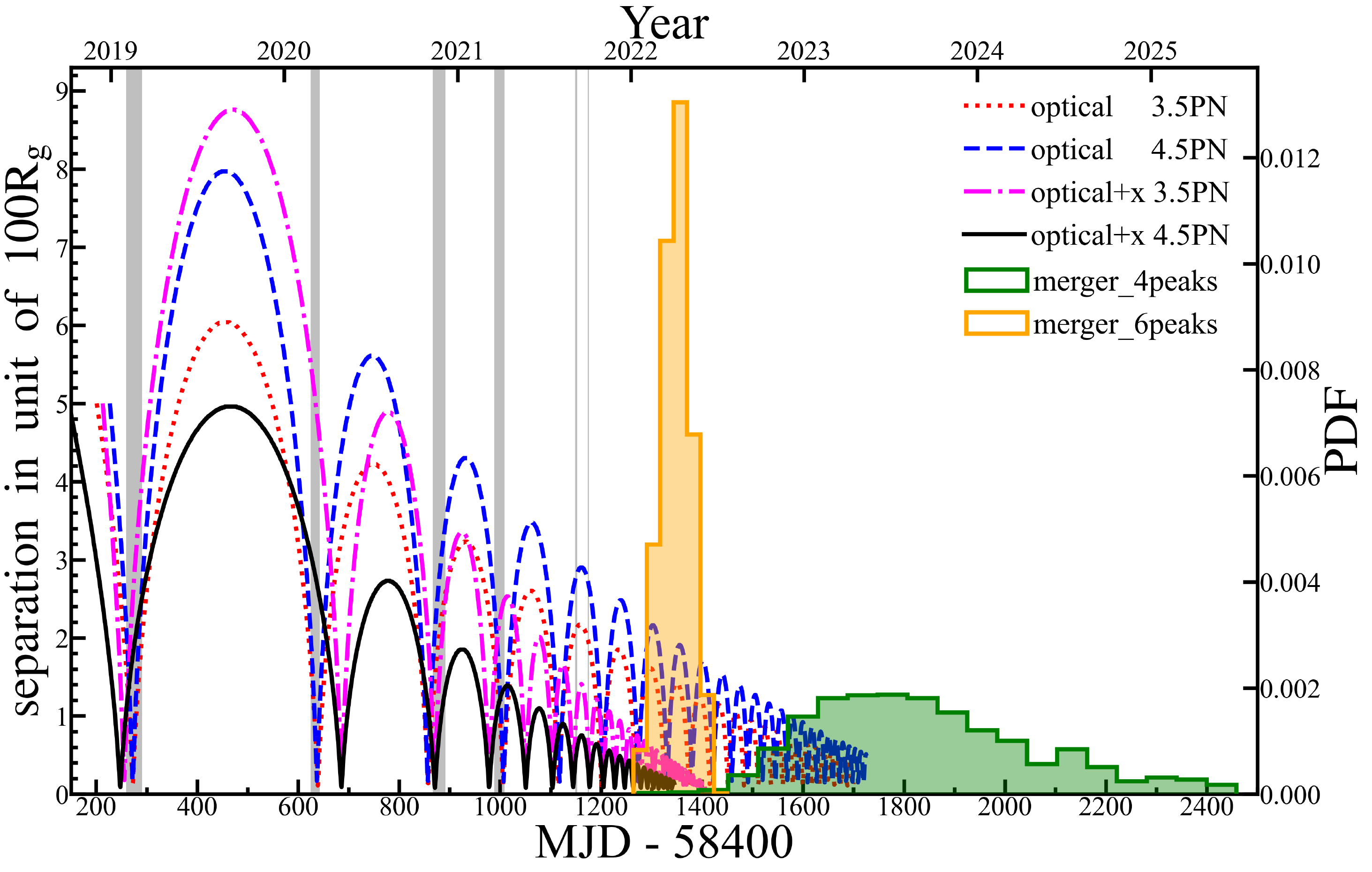}
\caption{
\textbf{The evolution of the binary separation from trajectory model and predicted merger time.}  The separations (in unit of $100R_g$ of the primary SMBH) inferred from different scenarios (with or without X-ray peaks, 3.5PN or 4.5PN, see details in Methods) are shown in red dotted, blue dashed, magenta dot-dashed and black lines, respectively. The observed peak times suggested by light curves are denoted with grey shadow regions. The possibility distribution function (PDF) of merger time (4.5PN) predicted from optical plus X-ray peaks is shown with orange histogram while that without X-ray peaks is shown in green histogram.
}
\label{lc2merg}
\end{figure}

 The host galaxy of SDSSJ1430+2303 appears as a typical elliptical galaxy with stellar mass of $1.5\times10^{11}$~\msun\ (see host galaxy part in Methods). In the standard framework of galaxy formation and evolution in a hierarchical universe, giant elliptical galaxies such as SDSSJ1430+2303 are considered to be the product of major mergers of two or more less massive galaxies. The delay time between galaxy merger and the final coalescence of SMBH binary could be as long as gigayears\cite{Binney1987}. This might explain why there is no visible morphological features indicative of galaxy mergers in the shallow SDSS image as the notable remnants have mostly dissipated.  On the other hand, 
 It is  interesting  to note that the SMBHB merger rate inferred by this event appears to be at least ten times higher than previous estimations (see related discussion in Methods).

 The expected  time till merger is approximately $100$-$300$ days considering both optical and X-ray light curves, and within three years considering only the optical data (see Figure~\ref{lc2merg}). Upon merger, the gravitational wave frequency is around the low-end of the frequency band of Laser Interferometer Space Antenna (LISA)\cite{amaro2017laser}, which is launching in 2030s, and well above the detection band of Pulsar Timing Array (PTA). The gravitational wave memory effect, however, may be detectable by PTA with additional five-ten years observation post merger. Although the black hole spins are unknown, because of the uneven mass ratio, the expected kick velocity is likely below $10^3$ km/s.
 
 Copious electromagnetic signals are expected \cite{Milosavljevic:2004cg}, from radio to X-ray band, binary inspiral stage to post merger, and from locations such as black hole corona, (possible) jet, accretion disk, interstellar medium, etc.  Neutrino production is also possibly detectable as a result of binary black hole coalescence \cite{Yuan:2020oqg}. We would like to call the attention of the astronomical community to perform extensive multi-messenger, multi-band observation on this transient AGN source. Hopefully this campaign will be celebrated by many unprecedented discoveries in the years to come.

\clearpage

\begin{methods}

{\bf Optical Light Curves} 

The core variability information, namely periodicity, of SDSSJ1430+2303 is revealed by ZTF as shown in Figure~\ref{lc}. The light curves are generated from ZTF forced PSF-fit photometry on difference images\cite{Masci2019} with reference flux added rather than photometry on original images because it allow more accurate variability measurement under different conditions (e.g., atmospheric seeing). We have first eliminated the problematic data by pixel quality flags and the low-quality data by setting a signal-to-noise ratio (SNR) threshold of 3.  Before ZTF, SDSSJ1430+2303 has been also occasionally covered by other surveys, including SDSS, CRTS\cite{Drake2009}, PanSTARRS\cite{Chambers2016} and ATLAS\cite{Tonry2018}. A combined analysis of all historic data is useful for characterizing its long-term behaviour and thus understanding the unique event  comprehensively. Except for ATLAS, which has scanned the sky with broadband filters, a "cyan" (c) band (420–650 nm) and an "orange" (o) band (560–820 nm), for asteroid search\cite{Tonry2018}, PanSTARRS and ZTF have used $g$ and $r$ filters in spite of slight bandpass differences. Ensuring a consistent photometry, we perform PSF photometry on the difference images by subtracting the ZTF references after zeropoint calibration. The cataloged CRTS data from aperture photomertry were simply used because of unavailability of their images. Finally, we obtain decade-long light curves of SDSSJ1430+2303 (see \extfig~\ref{oxlc}), which suggests that the central optical flux, measured by PSF photometry, began to rise since 2016 (or even earlier) and has brightened by $\sim0.7$ magnitude at $g$ band up to early 2018 (the initial stage of ZTF). The variability then switched to an unexpected periodic mode while its period and amplitude both show a uniformly decaying trend. 

{\bf  Mid-infrared Light Curves} 

SDSSJ1430+2303 has also shown a huge MIR flare revealed by its light curves from Near-Earth Object Wide-field Infrared Survey Explorer (NEOWISE)\cite{Mainzer2014}, which is a successor of WISE\cite{Wright2010}, and has thus been selected into the sample of MIR outbursts in nearby galaxies\cite{Jiang2021}. The NEOWISE all-sky survey has provided the photometry of SDSSJ1430+2303 at 3.4$\mu$m (W1 band) and 4.6$\mu$m (W2) with half-year cadence since late 2013. There are typically 12 individual exposures acquired within each epoch (within one day), which have been simply binned to improve their SNR. The MIR outburst started since 2017 January (MJD$\sim$57770) and reached to its peak around 2020 (see \extfig~\ref{oxlc}). It can be translated as the IR echo of the optical outburst, that is the reprocessed emission of dusty tori encircling SMBH, even its structure might have been disturbed dramatically by the other SMBH.

{\bf Swift Observation and Data Analysis}

We have requested a Swift ToO campaign to monitor SDSSJ1430+2303 since 2021 November 24 (PI: Jiang). A total of 18 snapshots have been obtained up to 2021 December 31, including 4 occasional visits before our program. We have downloaded the data from the HEASARC data center, and reprocessed them following the standard data reduction using tools in HEASOFT (v.6.29) with the latest calibrations. The event files have been rebuilt by task `xrtpipeline', with only the observations operated in `photon counting' mode being used. We extracted source photons within a circular region with radius of $47.2"$ and background photons from an annulus source-free region centered on our target. The net count rates in 0.3-10 keV band for each observation were 
then calculated (see detailed results in Extended Data Table~\ref*{xrt}) and result in its corresponding count rate light curve (see in Figure~\ref{lc}).

We have made a stacked spectrum by taking all the photon events together, in order to increase the low S/N ratio of each individual exposure, which possesses a total net exposure time of 19.06 ksec and an average count rate of $0.073\pm0.020$ cts/s. The new stacked spectrum were then regrouped, ensuring at least 20 counts per bin, for the purpose of $\chi^2$-statistics spectral fitting in XSPEC (v.12.12). The spectrum can be well fitted by a pure Galactic absorbed powerlaw with photon index of $\Gamma= 1.40\pm0.07~(\chi^2/dof=72.9/61; \rm N_{\rm H}=2.28\times10^{20}~\rm cm^{-2}$\cite{HI4PI}). The photon index is rather flat among AGNs. It is noticeable that the spectrum shows a smeared warm absorber feature around 0.7-1 keV (see \extfig~\ref{xspec}), which might affect the fitting. We then try to apply the Galactic absorption only model fitting to the 2-7 keV band and the photon index turns out to be more normal, namely $\Gamma=1.66\pm0.23~(\chi^2/dof=14.6/21)$). 
Aiming to get a more accurate description of the X-ray spectrum, we fix the photon index to $\Gamma=1.66$, yet adding an intrinsic ionized absorption component into the model. It yields an intrinsic column density of $N_{\rm H}=5.1_{-1.7}^{+2.4} \times10^{21}~\rm cm^{-2}$ and an ionization parameter of $\xi = 14.4_{-9.1}^{+20.3}~(\chi^2/dof=58.3/60)$. The best-fit unabsorbed flux given by the new fitting is $\rm 2.06\times, 3.15\times~and~5.21\times10^{-12}$~\ergs, that is $\rm 3.3\times, 5.1\times\ and~8.4\times10^{43}$~\lum\ in 0.3-2, 2-10 and 0.3-10 keV, respectively. The unabsorbed flux and luminosity of each individual observation can be easily derived from its corresponding count rate.

Besides XRT, SDSSJ1430+2303 has been also observed simultaneously by Ultraviolet Optical Telescope (UVOT)\cite{Roming2005} on Swift. However, the data taken before 2021 December 25 is too noisy due to insufficient exposure time in each filter. We then changed the observational strategy and focused on only UVW1 filter afterwards. The UVW1 photometry were measured with `uvotsource' task in the HEASOFT package using 4" apertures, giving AB magnitudes calibrated in the Swift photometric system\cite{Breeveld2011}. The UVW1 light curve show a similar peak around the second XRT peak.

{\bf Light Curve Peaks}

Four outstanding peaks appear in ZTF light curves up to 2021 August. Additional two peaks can be identified out in the subsequent Swift/XRT monitoring, which is about three months apart from ZTF.
In order to get a precise peak time, we try to fit the light curve profiles around each peak with Gaussian functions.  The uncertainties directly from fitting are probably underestimated because of simplified Gaussian assumptions. Thus we choose to put the uncertainties estimated from real data, that is the time interval when the magnitude has decreased by 1 sigma from peaks  (see results in Extended Data Table~\ref{peaktime}). Furthermore, the peak time differences between $g$ and $r$ bands are almost negligible taking into consideration of errors while $r$-band peaks show a tentative delay relative to that of $g$ band (by $\sim2.5$ days) if comparing their peak values straightforwardly.

{\bf XMM-Newton Observation and Data Analysis}

We have requested one XMM-Newton DDT observation on December 31 2021 (ObsID: 0893810201, proposer: Jiang) to get a high-quality spectra. We reprocessed the EPIC data reductions with the Science Analysis Software (SAS, v.20) and the latest calibration files. Only the data from the pn instrument of EPIC is used for our analysis, considering its better sensitivity than MOS. We created the events files with the tool of 'epchain'. After removing the `bad' pixels, we created the high flaring particle background time interval with a threshold rate of $>$0.6 cts/s with single events (`PATTERN==0') in 10-12 keV band. It results in a net exposure time of 38.56 ksec. We used only the single events for the science analysis. We extracted the source spectra from a circular region with radius of 32" centered on its optical position, and the background spectra from a nearby source-free circular region with radius of 40". We then regrouped the spectrum to have least 25 counts per bin to adopt the $\chi^2$-statistics spectral fitting in XSPEC.  We also created the background subtracted light-curves and corrected for various effects affecting the detection efficiency with the tool of 'epiclccorr'. We find that the count rates have increased slightly from 1.1 cts/s to 1.3 cts/s in 0.2-10~keV during the exposure. The average net count rate is $0.959\pm0.005, 0.232\pm0.003, 1.191\pm0.005$ cts/s in the 0.2-2, 2-10, and 0.2-10 keV bands, respectively. We find no obvious short quasi-periodic oscillation with careful power spectrum analysis. 
 
We have performed a simple Galactic absorbed powerlaw fitting to the pn spectra and obtained a photon index of $\Gamma=1.67\pm0.01$ ($\chi^2$/dof=792/616) . The warm absorber feature around 0.7-1 keV suggested by Swift/XRT spectra has been confirmed (see \extfig~\ref{xspec}). Then an intrinsic ionized absorption has been added into the model fitting ($\chi^2/dof=661.3/614)$), which results in an intrinsic column density of $N_{\rm H}=2.1_{-0.4}^{+0.6}\times10^{21}\rm cm^{-2}$, ionization parameter of $\xi= 29.9_{-14.1}^{+21.9}$, and a photon index of $\Gamma= 1.70\pm0.02$, that is also broadly consistent with Swift results. The unabsorbed flux in 0.3-10, 2-10 and 0.3-2~keV is $4.82\times, 2.83\times, \rm and~1.99\times10^{-12}$~\ergs, corresponding to $7.7\times, 4.5\times\rm and~3.2\times10^{43}$~\lum, respectively.
It is worthwhile to note that the absorber is better fitted by two components, one is outflowing and the other is inflowing with ultra-high velocity ($>0.1c$), while careful analysis and further observations are still necessary to confirm it. Such fast inflow has never been seen in other AGNs. It may  be associated with  disk crossings at small separations of the SMBH binary, which is another interesting topic that we will investigate in future works.

{\bf SDSS Spectral Analysis}

The optical spectra from Sloan Digital Sky Survey (SDSS), which is the  earliest one to our knowledge, was taken on 2005 May 4. After correction for Galactic extinction with the dust map\cite{Schlegeldustmap} and extinction curve\cite{Fitzpatrick1999}, we then fit its continuum with a combination of starlight and AGN power-law component (see details in \cite{Jiang2021}). 
Emission lines were left after removing the starlight and AGN continuum, which were then modelled with multiple Gaussians, including narrow (FWHM$<$800km/s) and broad (FWHM$>$1000km/s) components (see result in \extfig~\ref*{sdsspec}). 
A broad yet blueshifted (2390$\pm$174 km/s) \ha\ emission has been clearly detected. The absence of a similar component in \hb\ is probably caused by dust extinction.

In order to get the stellar velocity dispersion (\vdisp) of this galaxy, we first subtracted the fitted AGN continuum, resulting in a starlight dominated spectra. 
The \vdisp\ given by the fitting code pPXF\cite{Cappellari2004,Cappellari2017} is $\rm 182.3\pm8.4$~\kms, suggesting a \mbh\ of $2.1^{+2.0}_{-1.0}\times10^8$~\msun\ estimated from the empirical correlation between \mbh\ and \vdisp\cite{Kormendy2013}.
In contrast, the \mbh\ calculated by another empirical virial mass estimator for single-epoch AGN spectra \cite{GH2005} is $4.0^{+4.0}_{-2.0}\times10^7$~\msun\ given a \ha\ luminosity of $(1.80\pm0.09)\times10^{41}$~\lum\ and a FWHM of $6783\pm361$~\kms. The BLR size estimated from the radius-luminosity relation\cite{Bentz2013} is $\sim7$ light days, corresponding to $\sim1.5\times10^3$ Schwarzschild radius.

{\bf Optical Spectroscopic Follow-ups}

We have carried out a spectroscopic monitoring program of SDSSJ1430+2303 since late December in 2021. Five spectra have been obtained from Dec. 17 to Jan. 3 with YFOSC of the LiJiang 2.4m telescope (LJT for short) at Yunnan observatories and double spectrograph (DBSP) mounted on the Hale 200-inch telescope (P200 for short) at Palomar observatory (see observation details in Extended Data Table~\ref{spec_obs_info}). We reduced the YFOSC data with IRAF and DBSP data with Pypeit\cite{Pypeit(a),Pypeit(b)}, both of which provide standard procedures for reducing long-slit spectra. The new spectra is displayed in \extfig~\ref{1430_spec_follow} and compared with the SDSS spectrum. 

After correction for the Galactic extinction, we then modelled the continuum with a combination of starlight and reddened power-law, according to the same procedure of the continuum fitting of SDSS, while the starlight shape of LJT spectra were fixed to that from DBSP due to lower S/N. Aiming to inspect the change of \ha\ and \hb\ profiles clearly, we first normalized their flux to the one from DBSP, which has the highest S/N. On the other hand, the flux of DBSP spectrum itself has been scaled to ensure an equivalent $\rm [O\,\textsc{iii}]\lambda5007$ flux to SDSS spectra since the emission from AGN narrow-line region should keep almost constant within decade timescale. The emission line profiles around \ha\ and\hb\ regions have exhibited remarkable change with respect to SDSS spectrum (see right panel of \extfig~\ref{1430_spec_follow}) although no rapid variations are detected between the four follow-up spectra. Most notably, an obvious broad redshifted component appeared. We tried to fit the broad \ha\ and \hb\ emissions with mutliple-Gaussians, during which an additional Gaussian is considered as necessary only when it can significantly improve the fitting evaluated by $F$ test. The final fitting demands three broad Gaussians, including an obviously blueshifted ($\sim4000$~\kms) and redshifted one ($\sim4600$~kms), respectively.

{\bf Host Galaxy properties}

SDSSJ1430+2303 appears as a typical red elliptical galaxy with no apparent signatures of galaxy interaction in SDSS images (see the left panel of \extfig~\ref{sed}). In order to acquire the galaxy properties quantitatively, we have gathered its multiwavelength photometry  spanning from ultraviolet to mid-infrared and performed a fitting of the spectral energy distribution (SED). The Code Investigating GALaxy Emission (CIGALE)\cite{Boquien2019} has been chosen by us to model the SED since it has considered stellar, dust and AGN components reasonably. During our fitting, we have simply used a delayed star formation history assuming a single starburst with an exponential decay. A single stellar population (SSP)\cite{Bruzual2003}, default nebular emission model\cite{Inoue2011} and dust attenuation module\cite{Calzetti2000} have been employed. In addition, we utilize the model provided by Draine\cite{Draine2014} to account for dust emission while the model of Fritz\cite{Fritz2006} for AGN emission. 

The fitting works well ($\chi_{\nu}^2$=2.1, see the result in the right panel of \extfig~\ref{sed}), giving a stellar mass of $1.52\times10^{11}$~\msun\ dominated by old-aged stellar population ($\sim$12~Gyr).  The AGN has contributed only 6\% to the total emission of this galaxy. The acquired stellar mass predicts a \mbh\ of $\sim8\times10^8$~\msun\ if naively assuming that the whole galaxy is classical bulges\cite{Kormendy2013}.

{\bf Alternative Models}

Besides binary SMBHs, it is instructive to examine other  scenarios that could give rise to chirping AGN flares. While it is difficult to consider all viable cases, we shall discuss three scenarios that are natural to consider.

In the first scenario, the companion is a star or a dense gas clump which is decelerated every time it hits the accretion disk. Consider a standard thin disk model, the surface density is
\begin{align} \label{eq:disk2}
    \Sigma(r) \approx
    \begin{cases}
   1.7\times 10^4 {\rm g \,cm^{-2}} (0.1/\alpha) (0.1\dot{M}_{\rm Edd}/\dot{M})(r/10^2M)^{3/2} \quad \text{for $\alpha$-disks}\,,\\
   1.45\times 10^6 {\rm g \,cm^{-2}} (0.1/\alpha)^{0.8} (\dot{M}/0.1\dot{M}_{\rm Edd})^{0.6}(M/2\times 10^8 M_\odot)^{0.2}(10^2M/r)^{0.6}  \quad \text{for $\beta$-disks}\,,
   \end{cases}
\end{align}
where the $\alpha$-disk has a viscous stress $t_{r\phi}=-1.5 \alpha p_{\rm tot}$, with $p_{\rm tot}$ being the total pressure in the disk, and the $\beta$ -disk model assumes $t_{r\phi}=-1.5 \alpha p_{\rm gas}$, with $p_{\rm gas}$ being the gas pressure. It is straightforward to check that the work done by dynamical friction, for a stellar -mass companion, is
\begin{align}\label{eq:friction_force}
W_{\rm DF} \approx\frac{1}{v^2}4\pi M_\odot^2\Sigma C\,
\end{align}
which turns out to be much smaller than the kinetic energy $0.5 M_\odot v^2$ (with $C \sim \mathcal{O}(1)-\mathcal{O}(10)$). The most dominant effect comes from the head wind, with which the companion captures gas up to a fraction of its mass so that its orbit can be significantly affected. Let us denote the size of the companion to be $R_c$, requiring $\pi R^2_c \Sigma \sim M_\odot$ leads to $R_c \sim 2 \times10^8 $~km for $\beta$-disk at $r \sim 10^2 G M/c^2$ and $2\times 10^9$~km for $\alpha$-disk, which are much greater than the typical size of a star. Notice that gas clump with such radius has gravitational binding energy $M_\odot/R_c$ which is much less than the amount of kinetic energy converted to heat after each collision with the disk, i.e., the clump would disrupt. Therefore this star/gas clump scenario is not likely the origin of the discovered chirping AGN. 

In the second proposal a star loses its orbital energy as the internal oscillation is excited during the pericenter passage, which subsequently converts into heat. The specific orbital energy loss is $\sim G M/r \sim \mathcal{O}(1/500)$, which is much larger than the specific gravitational binding energy of a normal star $\sim G M_\odot/R_\odot \sim 2\times 10^{-6}$. So such a star would disrupt before being able to finish the next pericenter passage. One may instead imagine that a star partially disrupts and releases a small fraction of its mass during the pericenter passages to power the flares. In this case the pericenter distance has to be close to the tidal disruption radius. Besides the excitation of internal oscillations,   it is straightforward to show that the gravitational wave radiation  is also insufficient to drive the orbital decay, because of the extreme mass ratio ratio. Therefore it is still incompatible with the fast changing periods. Along a similar line, a stellar-mass black hole binary may serve as the companion with its center-of-mass orbital energy transferring to its internal energy by tidal gravitational coupling. The transfer efficiency is maximized when the internal orbital frequency is comparable to the pericenter frequency, and rapid orbital decay is possible without the need of considering gravitational wave radiation. However, an energy budget calculation using the dynamical friction shows that stellar-mass black holes are unlikely able to generate the observed $>\mathcal{O}(10^{50})$ergs flare as they collide with the disk.

The third  possibility is related to disk activities, i.e., oscillations and/or instabilities. For example, disk instabilities have been proposed as possible explanations of quasi-periodic eruptions (QPEs) \cite{Miniutti:2019fqr}. Quasi-periodic oscillations (QPOs) are also found in SMBH systems \cite{gierlinski2008periodicity}. In addition to the difference in the frequency band of detection, the chirping flares we observed are distinct from QPOs and QPEs in at least two major aspects. First of all, the characteristic radius inferred from the first flare is likely greater than $500$ gravitational radius, which is $\mathcal{O}(10)-\mathcal{O}(10^2)$ times of those inferred from QPEs and QPOs. Secondly, the fast varying period is not seen in QPEs and QPOs. In fact, if we convert the period into radius using the Kepler's law, and consider the time dependence of this radius, the speed it shrinks is significantly different from the local sound speeds. it is unlikely that the chirping AGN is primarily contributed by known disk activities. 

{\bf Trajectory Model}

We consider a  black hole binary following a highly eccentric orbit. Such systems, in the stellar-mass range, have been one of the poorly modelled scenarios for ground-based gravitational wave detectors. A common technique for modelling the orbital evolution is to apply the Post-Newtonian (PN) formalism, which effectively expands the relativistic equation of motion in $1/c$ and $G$. As the Post-Newtonian expansion is only known to a certain order, using a truncated PN equation of motion inevitably brings in theoretical errors. Such errors are expected to be more important when two black holes are close to each other, i.e., in the late inspiral stage when higher PN corrections grow up. Highly eccentric binaries are more susceptible to such errors, as a small uncertainty of the description near the pericenter tends to be amplified over the long orbital period, so that the timing of the next pericenter passage is significantly influenced. For the construction of gravitational waveform model, the phase error is generally required to be within $\mathcal{O}(0.1)$ level (so the timing accuracy should be better than $\mathcal{O}(0.1)$ of the pericenter passage timescale), which is beyond the capability of current PN waveform models for highly eccentric binaries. In recent years there are developments using the Effective-One-Body formalism, but the waveforms are mostly calibrated for eccentricity $e \le 0.3$ \cite{Nagar:2021gss}, which is clearly not applicable for highly eccentric binaries with $e>0.9$.

In the case we study the measurement uncertainty of the peak time of the flares is around $20-30$ days, which is much longer than the pericenter passage timescale ( less than a day). Therefore our requirement on the timing accuracy of the trajectory model is less stringent than that for constructing gravitational wavveforms. However, it is worth to note that the error may accumulate over multiple pericenter passages to cause larger mismatch at later times. In our implementation we truncate the equation of motion at different PN orders, and compare results to estimate the systematic error brought by the PN formalism.

We apply the PN equation of motion for the two-body system written in the Modified Harmonic gauge \cite{Blanchet:2013haa}.
\begin{align}
 \frac{d {\bf v}}{d t} =& -\frac{G M}{r^2} \left[ \left( 1+ \mathcal{A} \right) {\bf n}+ \mathcal{B} {\bf v} \right] \nonumber \\
	&- \frac85 \frac{\nu}r \left( \frac{G M}{r} \right)^2\left( \dot{r} \mathcal {A}_{RR} {\bf n} + \mathcal{B}_{RR} {\bf v} \right),
\end{align} 
where $M$ is the total mass, $\nu\equiv M_1 M_2/M^2$, ${\bf v}$ is the relative velocity and $\bf n$={\bf r}/r is the normalized relative position vector. The PN coefficients $\mathcal{A}, \mathcal{B}, \mathcal{A}_{\rm RR}, \mathcal{B}_{\rm RR}$ can be found in ref.\cite{Blanchet:2013haa} and we have kept them up to $4.5$PN order. The tail term showing up at $4$PN order is nonlocal in time, which is not included here.
In general the spins of black holes enter the equation of motion starting at $1$PN. As the information from flare timing is limited, we neglect the spin degrees of freedom to simplify the trajectory model.

%

With the trajectory model we are ready to check how well it fits the observed optical and X-ray light curves (see \extfig*\ref{lcfit}). We shall identify the time of periceter passage as approximately the time of the peak luminosity, as in most cases the secondary black hole crosses the disk shortly before and after the pericenter passage. The peak times shown in Extended Data Table.~\ref{peaktime} are obtained using Gaussian profiles to fit the data around the peaks of the flares, and the peak-time uncertainties are estimated by demanding that the drop in flare luminosities exceed the uncertainties of the peak values. We define the Likelihood function to be
\begin{align}\label{like}
  \log  \mathcal{L} \propto - \sum_i \frac{(t_i-T_i)^2}{2\sigma^2_i}\,,
\end{align}
with $t_i$ being the ith pericenter passage time predicted by the trajectory model, $T_i, \sigma_i$ are the ith  peak time and uncertainty as shown in Extended Data Table~\ref{peaktime}. In addition, we have excluded the cases where additional peaks are predicted near the trough regions of the light curves.

We have performed Markov-Chain Monte-Carlo simulations based on the likelihood function in Eq.~\ref{like}.
The four parameters contained in the trajectory model are $M_1, M_2, E_0, p$, where $E$ is the initial orbital energy (as normalized by $G M/r_0, r_0= 500 G M_1/c^2$), and $p$ is defined as $L_0 :=M_1 M_2/M \sqrt{2 M p}$, with $L_0$ being the initial angular momentum. Physically $p$ can be viewed as the initial pericenter distance in the Newtonian regime, although it may differ significantly from the actual pericenter distances computed using the PN trajectory model.

The posterior distributions of $M_1, M_2, E_0, p$ are shown in \extfig~\ref{cornerp1} and \ref{cornerp2} with respect to four different settings respectively:  Optical$+$X-ray$+3.5$ PN, Optical$+$X-ray$+4.5$ PN, Optical$+3.5$ PN, Optical$+4.5$ PN. The corresponding trajectories for the maximum likelihood sample point in each cases is shown in \extfig~\ref{lcfit}.  In the latter two cases the X-ray peaks are singled out as they are expected to come from the corona of the disk, which may be susceptible to the accretion conditions of the inner disk and/or dynamical variations of the corona itself. 
The four-peak cases generally fit data better than the six-peak cases - the ratio between the maximum likelihood in these two cases is approximately $400$. This is also illustrated in the posterior distribution plot shown in \extfig~\ref{cornerp1} and \ref{cornerp2}, where we find the four-peak and six-peak fittings prefer rather distinct regimes of $E_0 - p$ parameter space. In particular, the six-peak Monte-Carlo simulation suggests that most of the $E_0$ distribution is above zero, i.e., the orbit is unbound, where the four-peak samples allow significantly more weight in the bound case. Physically both bound and unbound orbits are viable possibilities, but the unbound case seems to be more difficult to interpret from the formation channel point of view (see the discussion in the next section). The simulation  assuming $4.5$PN equation of motion gives rise to similar results.

The distribution of expected merger time can be computed by evolving the Monte-Carlo samples till their mergers. The posterior distributions for the four cases are shown in \extfig*\ref{mtimepos}. In general the six-peak cases predict $100$-$300$ days till merger, counting from the second peak time in X-ray, whereas the  four-peak case constrain the merger time to be within $\sim$ three years.

We have adopted a rather simplified trajectory model to control the number of model parameters, as compared with the previous exercise on OJ-287 \cite{Dey:2019pwb}, because of limited number of flares observed so far. In the future when more peak timing information is available, the model can be improved in several aspects. First, as we mentioned earlier, the spins of black holes also enter the equation of motion, so it will be beneficial to include the spins, at least for the primary black hole. Second, the pericenter distances inferred from our model fitting are $\sim 10 G M/c^2$. At such close distances Post-Newtonian theory may not be the most accurate way of describing the orbital evolution. It might be possible to combine the black hole perturbation technique, and design a hybrid model (PN at far distance $+$ black hole perturbation near pericenter passages) that better deal with the motion in the strong-gravity regime. Third, strictly speaking, there is a time delay between the disk-crossing time and the peak of the flare, which is approximately $t_{\rm d} = C_{\rm d} M^{1.24}_2 v^{-4.23}_{\rm rel} H^{-0.29} \Sigma^{0.91}$, where $v_{\rm rel}$ is the relative velocity, $H$ is the disk height and $C_{\rm d}$ is a parameter to be determined from the data. We have neglected this piece as the inferred pericenter distances are all close to $10 GM/c^2$ for the first several peaks, so the delay times may not differ significantly from each other for these  flares (see the next section for details). A constant delay time is naturally accounted for as we freely choose the starting time of the trajectory model. In future implementation  as the three-dimensional trajectory prescription is used, we may compute $t_d$ based on the flare model for each disk-crossing and include them into the trajectory model. 

{\bf Flare Model}

As a black hole collides with an accretion disk, it kicks out certain amount of gas within radius $R_a = C_{\rm BH} G M_{\rm BH}/v^2_{\rm rel}$ ($C_{\rm BH}$ is a dimensionless constant of order unity) which later on expands and radiates when the plasma ball becomes optically thin. This plasma ball model has been used to explain the optical flares of OJ-287 \cite{Lehto1996,Pihajoki2016}. We shall apply the same model to analyze the characteristics of optical flares discussed in this work. However, it is also worth to note that disk heating due to the collision shocks, either the  ones coming from black hole-disk crossings or the colliding accretion flows originally moving around individual black holes, may also radiate in the optical band. Indeed it is possible that the first optical flare is contributed by multiple mechanisms.

The energy associated with the peaks of the subsequent flares (in g/r bands), are approximately $\mathcal{O}(10^{50})$ ergs. The characteristic decay timescales of the luminosity function after each peak are listed in Table.~\ref{peaktime}. These numbers are broadly consistent with an $\alpha$ disk with profile parameterized in Eq.~(\ref{eq:disk2}).
For each collision of the secondary BH on the disk, a total amount of energy $\delta E = \frac{1}{2} \Sigma \pi R_a^2 v_{\rm rel}^2$ is deposited into the shocked gas which subsequently eject from the disk, with
\begin{equation}
    \delta E =2.5\times 10^{50}\ {\rm ergs}\ C_{\rm BH}^2  \left (\frac{0.3}{\alpha}\right )
\left (\frac{0.1\dot M_{\rm Edd}}{\dot M} \right ) \left (\frac{M_2}{4\times 10^7 M_\odot} \right )^2 \left (\frac{r}{20 M} \right )^{2.5}
\delta^{-2}\ ,
\end{equation} 
for an $\alpha$ disk, 
where we have written the relative velocity in terms of the local Keplerian velocity as $v_{\rm rel}=\sqrt{2} v_{\rm K} \delta$, with $\delta =\sqrt{1-\sin \iota}$ being a geometrical factor ($\iota$ is the angle between the velocity of the black hole and the normal direction of the disk). 
For each collision, the delay time $t_{\rm d}\propto r^{3.48}\delta^{-4.23}$, the emission duration $t_{\rm b}\propto r^{1.29}\delta^{-1.40}$,
and 
the ratio of the two \cite{Pihajoki2016} 
\begin{align}\label{eq:t_db}
t_{\rm d}/t_{\rm b}&=0.019 \left ( \frac{H}{10^{15} \ {\rm cm}}\right )^{0.14} \left (\frac{n}{10^{14}\ {\rm cm}^{-3}}\right )^{0.52} \left (\frac{M_2}{10^8 M_\odot} \right )^{0.29} \left (\frac{r}{20 M} \right )^{1.41} \delta^{-2.82}\nonumber \\
& =1.6\times10^{-3} \left (\frac{\alpha}{0.3}\right )^{-0.52} \left (\frac{\dot M}{0.1 \dot M_{\rm Edd}}\right )^{-0.89} \left (\frac{M}{2\times10^8 M_\odot} \right )^{-0.38}  \left (\frac{M_2}{4\times10^7 M_\odot} \right )^{0.29} \left (\frac{r}{20 M} \right )^{2.19} \delta^{-2.82}\ .
\end{align}
From Table.~\ref{peaktime}, we find small variation in the emission timescale $t_{{\rm b}, i} (i=1, 2, 3, 4)$ from collision to collision, while the peak energy $\delta E_i$ decreases with time. Making use of the  scaling relations for $\delta E$ and $t_{\rm b}$, 
we infer  the ratio of each collision radius $r_j$, geometrical factor $\delta_j$ 
and delay times $t_{{\rm d}, j}$ ($j=2, 3, 4$) as 
\begin{equation}
     r_1/r_j = (1.2, 2.6, 3.7)\ , \quad \delta_1/\delta_j = (1.0, 2.5, 3.5)\ ,  \quad t_{{\rm d}, 1}/t_{{\rm d}, j} = (1.5, 0.6, 0.5)\ .
\end{equation}
based on the g-band data. 
The delay times of the latter three collisions differ $\sim 50\%$ from the first one, and in general the delay time is much shorter than the emission duration (Eq.~(\ref{eq:t_db})). Therefore the plasma ball model suggests that neglecting the delay times in fitting the trajectory model to flaring peaks introduces negligible bias.
The flare model correctly predicts that the collisions happen closer and closer to the primary SMBH, however the inferred collision radii  should be subject to model uncertainties which could be calibrated by a 3d trajectory model if more emission peaks are measured by subsequent observations.

{\bf Event Rate}

In the standard paradigm of binary SMBH formation, after the merger of two galaxies, 
the separation between black holes initially increase because of the scattering with stars, dynamical friction and possible interaction with environmental gas. When the separation is below $\mathcal{O}(0.1)$pc scale, the gravitational wave radiation takes over as the main dissipation mechanism all the way towards merger. In most of the previous rate calculations for SMBH mergers,
ultra-high eccentricity ($e>0.95$) binary or hyperbolic encounter have never been considered as important intermediate stages before the final binary coalescence. This type of system  is interesting as the close pericenter passages greatly accelerate the binary merger process in the final stage.

Theoretically it has been pointed out that such high eccentricity binaries may be produced through scattering with chaotic field stars, as their angular momentum are randomized by non-axisymmetric potential due to the Kozai mechanism \cite{Iwasawa:2010qr}. A hierarchical SMBH triple system may also achieve high eccentricity for the inner binary through the Kozai mechanism. It  however remains unclear how effective these mechanisms apply for general binary mass ratios, and how much they affect the rate of SMBH mergers. The Monte-Carlo simulation from previous section shows a significant chance that the binary forms from an unbound system. It is also unclear whether this scenario can be interpreted using similar mechanisms.

We may estimate the event rate for similar type of AGNs with varying periods as follows.
In a Poisson process with rate $\lambda$, if we only detect one signal up to time $t_0$, the probability is $P(N(t\le t_0)=1 | \lambda) =\lambda t_0 e^{-\lambda t_0}$. This relation can be inverted according to the Bayes theorem, $P(\lambda | N(t\le t_0)=1) \propto P(N(t\le t_0)=1 | \lambda)$, to obtain the posterior distribution for $\lambda$, for which we estimate the $1\sigma$ range to be $\in [1.15, 2.35]t^{-1}_0$. Similar argument applies for the searched (higher-dimensional) volume v.s. number of detected events.  The redshift of SDSSJ1430+2303 is 0.08105 and the duration of light curve is about 4 years (until the expected merger). However, we have also searched for all known SDSS AGNs within $z\le 0.35$\cite{Liu2019} for similar period of time but have not found a second chirping source. As a result, the inferred event rate is between $(2.4-4.8) \times 10^{-2}$ ${\rm Gpc}^{-3} {\rm yr}^{-1}$. This number may also change as we search over AGN samples in the deeper universe. Note this rate is on the high end as compared with the SMBH binary merger rate and galaxy merger rate, which are estimated to be below $2 \times 10^{-2}$ ${\rm Gpc}^{-3} {\rm yr}^{-1}$ \cite{Klein:2015hvg}. The discrepancy may be amplified if we consider the fact that generally only $\sim 1\%$ local galaxies host AGNs\cite{Mo2010} while the fraction among merging galaxies is $\sim10$ times higher\cite{Weigel2018}, so that every chirping AGN approximately has ten invisible partners at the similar stage of SMBH binary merger. On the other hand, it is reported that in the dwarf galaxy Leo I a SMBH with comparable mass to Sgr A* is found \cite{Bustamante-Rosell:2021ldj}, despite the fact that Leo I is $\mathcal{O}(10^{-5})$ times lighter than the Milky Way. In this sense minor mergers with extreme mass ratios should be included in the galaxy merger tree calculations, as the associated mass ratio of the SMBH binary may be less uneven. Therefore it is possible that the true rate of SMBH mergers is indeed ten times higher than the previous estimations.

{\bf Gravitational Wave Detection}

The expected merger time of this SMBH binary is within one to three years, so this event will be missed by LISA. On the other hand, The inspiral and merger frequency of the binary is several orders of magnitude higher than the frequency band of Pulsar Timing Array (PTA), so we will not be able to probe the oscillatory part of the gravitational wave signal. The gravitational memory piece of the waveform, however, scales as $f^{-1}$ at low frequencies which is possibly observable by PTA \cite{vanHaasteren:2009fy,Pshirkov:2009ak}. We now present a brief calculation to address the expected signal-to-noise ratio (SNR).

The SNR is given by $h/\sigma$. Here the r.m.s (root-mean-square) noise $\sigma$ depends on the number of pulsars using, the pulsar timing residuals, the observation duration, etc. We adopt a simple model used in \cite{vanHaasteren:2009fy}, by assuming the number of working pulsar is twenty, the average r.m.s residuals of these pulsars is on the level of $100$ns, the measurements are performed biweekly, and the observation duration post merger is five years (pre merger observation assumed to be five years as well). The noise is assumed to be whitened after removing the red noise and the corresponding $\sigma$ in this idealized scenario is $\approx 4.5 \times 10^{-16}$. This number should be updated considering the sky location, number and quality of pulsars available from the International Pulsar Timing Array Collaboration. The magnitude of the memory $h$, on the other hand, is given by \cite{Favata:2009ii}
\begin{align}
    h = \frac{1}{384 \pi d} \sin^2\Theta (17+\cos^2\Theta) \int^{T_{\rm R}}_{-\infty} (I^{(3)}_{22})^2 dt\,,
\end{align}
where $d$ is distance to the source, $\Theta$ is the inclination angle of earth in the source frame, $T_{\rm R}$ is the duration of observation post merger, and $I^{(3)}_{22}$ is the third time derivative of the $\ell=2,m=2$ mass quadrupole moment of the binary. Note that we use this formula with $I_{22}$ derived for circular binaries as an estimation for $h$ because it is sensitive to the total gravitational wave energy radiated instead of the phase of the waveform. In addition, its main contribution comes from cycles before and after the merger, and that eccentric binaries circularize quickly in the late inspiral stage.

We use the Monte-Carlo samples obtained from fitting the trajectory model with the light curves to extract the black hole masses, and sample $\Theta$ assuming a uniform distribution in the sky of the source. The quadrupole moment $I_{22}$ can be approximately computed using the expression in \cite{Favata:2009ii}. The corresponding distribution of SNR is shown in \extfig~\ref{fig:snr}. For a duration of five years observation post merger, the expected SNR is around one, or approximately two in the best-case scenario.
Note this number can only serve as a guide as the estimation for $\sigma$ is very crude. In reality some of best pulsars are observed weekly, and the available data sets span for more than five years. It will be important to take into account all the info prior to the merger and combine them with new data post-merger to obtain more realistic SNR.

{\bf Electromagnetic Signal Upon Merger}

The electromagnetic counterparts associated with SMBH binaries have been extensively studied in the past. One thing to note is that the expected binary orbit should be inclined with respect to a geometrically thin accretion disk, which is different from the in-plane assumption made by most available 2D and 3D GRMHD simulations.  The dissipation timescales for the flare energy, as estimated from the optical light curve, range between 25 days to 35 days, which are longer than the orbital periods in the late stage of binary inspiral. So the resolvable oscillations in optical luminosity may gradually fade away  before merger. However, the background floor of the optical light curve may continuously rise towards merger because of the repeatedly heating from the shock waves generated by black hole collisions with the disk.

The viscous timescale for a standard thin disk is
\begin{align}
   t_{\rm vis} \approx
    \begin{cases}
   4\times 10^7  (0.1/\alpha) (0.1\dot{M}_{\rm Edd}/\dot{M})^2 (M/2\times 10^8 M_\odot) (r/10M)^{3.5} \,{\rm secs}\quad \text{for $\alpha$-disks}\,,\\
   5\times 10^9  (0.1/\alpha)^{0.8} (0.1\dot{M}_{\rm Edd}/\dot{M})^{0.4}(M/2\times 10^8 M_\odot)^{0.8}(r/10M)^{1.4}  \,{\rm secs}\quad \text{for $\beta$-disks}\,,
   \end{cases}
\end{align}
which is likely longer than the remaining merger time, so the part of disk within the black hole crossing radius will not reach equilibrium prior to merger. In fact, each black hole-disk crossing at radius $r$ approximately transfers $\Sigma(r)\pi  \Delta r^2 v^2$ energy to the gas, with $\Delta r \sim G M_2/v^2$. As $v \sim G M/r$,  we find that a fraction of $(M_2/M)^2$ gas within radius $r$ is significantly affected. Because there are several tens of cycles in total before the final merger, the gas within this inner region gains additional energy comparable to the original Keplerian energy. The inner disk will gradually become ``puffed-up", i.e., no longer being geometrically thin. Therefore we may not be able to observe the iron K$\alpha$ emission line from the X-ray spectroscopy measurement. It is also unclear whether and how these energies may feed into the hot corona around the supermassive balck hole(s).


The X-ray emission, expected from the corona of the SMBH(s), may comprise luminosity variation in multiple frequencies. The in-plane, circular binary GRMHD simulations typically suggest at least two frequency components associated with the radial oscillation of a lump of mass and accretion events as black hole mini-disks pulling matter from the lump \cite{Noble:2021vfg,Gutierrez:2021png}. 
An inclined, eccentric and uneven mass-ratio binary likely produces more than two harmonics, with frequencies changing in time. It will be interesting to Fourier decompose the time-domain X-ray light curves to identify the physical origins of various Fourier components.

With the Monte-Carlo samples from the trajectory fittings, we also sample dimensionless spins with uniform prior from $0$ to $1$ and isotropic distribution in their directional vectors. The resulting kick velocity is mostly below $10^3 {\rm km s^{-1}}$ \cite{Varma:2018aht}, as shown in \extfig~\ref{fig:kick}.  The final black hole carries accretion gas within radius $\sim G M/v^2$, which is larger than $10^5$ gravitational radii. The infall of gas towards the final black holes takes years based on the viscous timescale. 
The final black hole will also carry along the orbiting clouds, so that we may expect the projected velocity of BLR clouds to be shifted again, but only for large enough kick velocities.   
A jet may be launched after the accretion flows onto the final black hole gradually settles, which possibly produces observable neutrino emissions \cite{Yuan:2020oqg}.

\clearpage

\begin{figure*}
\renewcommand{\figurename}{Extended Data Fig.}
\setcounter{figure}{0} 
\centering
\begin{minipage}{1\textwidth}
\centering{\includegraphics[angle=0,width=1.0\linewidth]{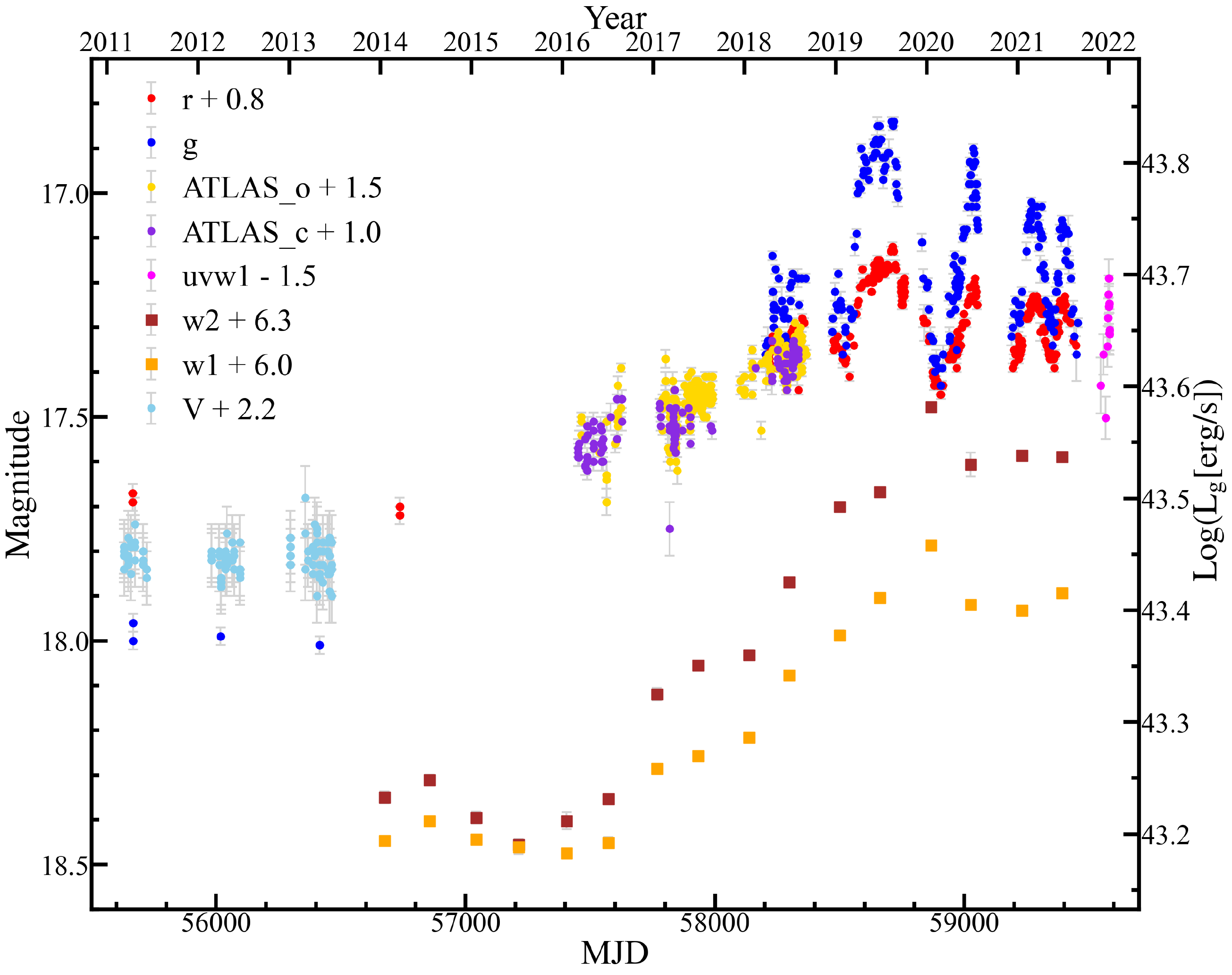}}
\end{minipage}
\caption{
\textbf{The long-term optical light curves of SDSSJ1430+2303}. We have gathered useful photometric data in the past decade, spanning from PanSTARRS (2011-2014, g band in blue and r band in red), CRTS (2011-2013, V band in skyblue), ATLAS (2016-2018, c band in blueviolet and o band in gold) to ZTF (2018-2021). All photometry except for CRTS have been calibrated to ZTF reference flux with PSF photometry. The g band monochromatic luminosity ($\nu L_{\nu}$) has been also denoted in the vertical axis on the right side. The NEOWISE W1 and W2 data are shown as orange and blue squares, respectively.
}
\label{oxlc}
\end{figure*}

\clearpage

\begin{figure*}
\renewcommand{\figurename}{Extended Data Fig.}
\centering
\begin{minipage}{1\textwidth}
\centering{\includegraphics[angle=0,width=1.0\linewidth]{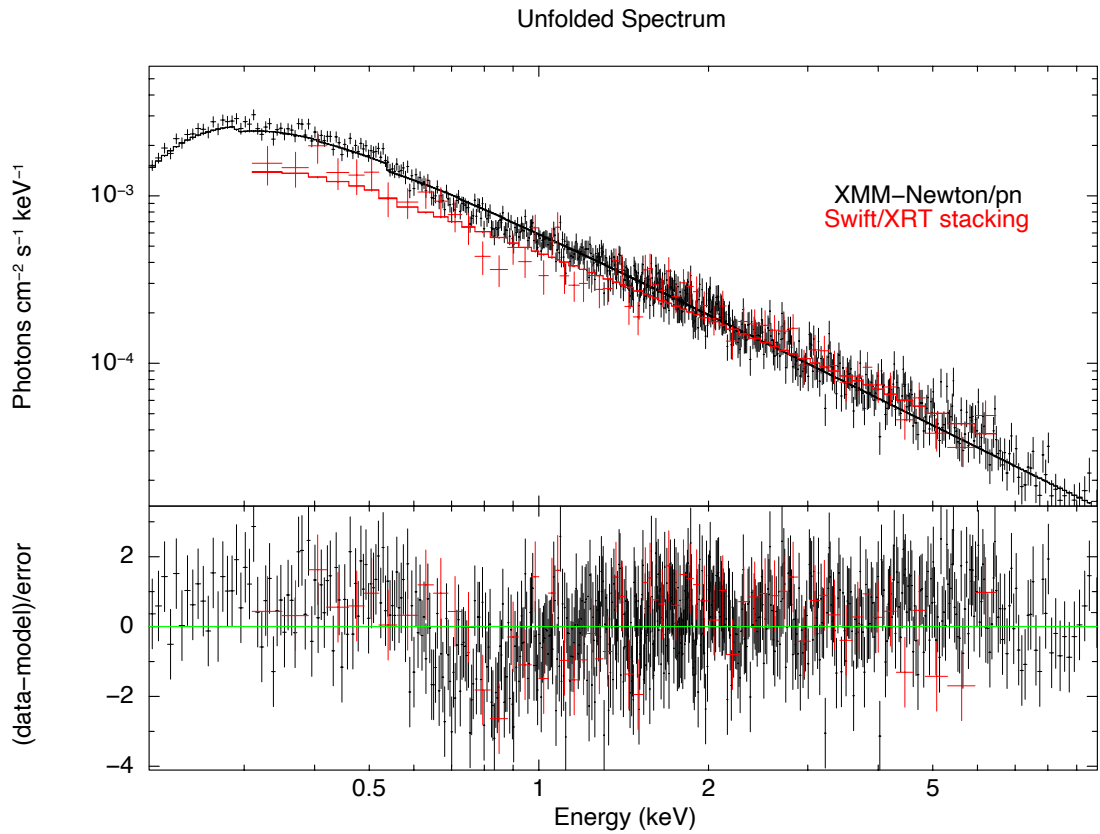}}
\end{minipage}
\caption{
\textbf{XMM-Newton (black) and Swift/XRT stacked (red) X-ray spectra.}. 
The model fittings have been overplotted in black and red lines, respectively. An obvious absorber feature appears between 0.6-0.9~keV in both residuals.
}
\label{xspec}
\end{figure*}

\clearpage

\begin{figure*}
\renewcommand{\figurename}{Extended Data Fig.}
\centering
\begin{minipage}{1\textwidth}
\centering{\includegraphics[angle=0,width=1.0\linewidth]{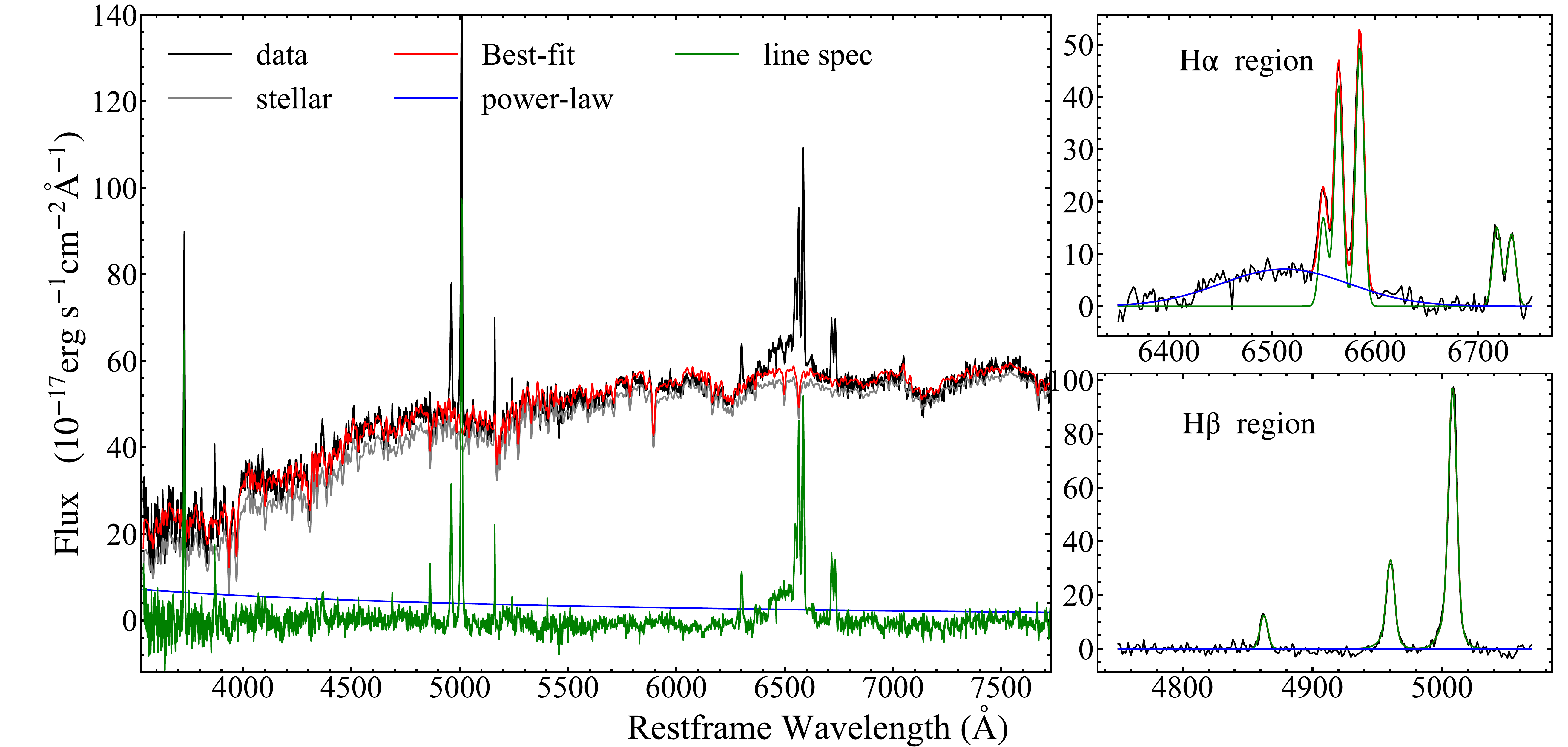}}
\end{minipage}
\caption{
\textbf{The SDSS spectrum and its decomposition of SDSSJ1430+2303.} The left panel displays the decomposition of the whole spectra, in which the starlight, AGN continuum and their sum shown in grey blue and red, respectively. The continuum subtracted residual is plotted in green. The right top panel shows the Gaussian fitting to the \ha\ region, in which the broad, narrow, and total emissions are presented in blue, green, and red, respectively. The right bottom panel is similar but for \hb\ region.}
\label{sdsspec}
\end{figure*}

\clearpage

\begin{figure*}
\renewcommand{\figurename}{Extended Data Fig.}
\centering
\begin{minipage}{1\textwidth}
\centering{\includegraphics[angle=0,width=1.0\linewidth]{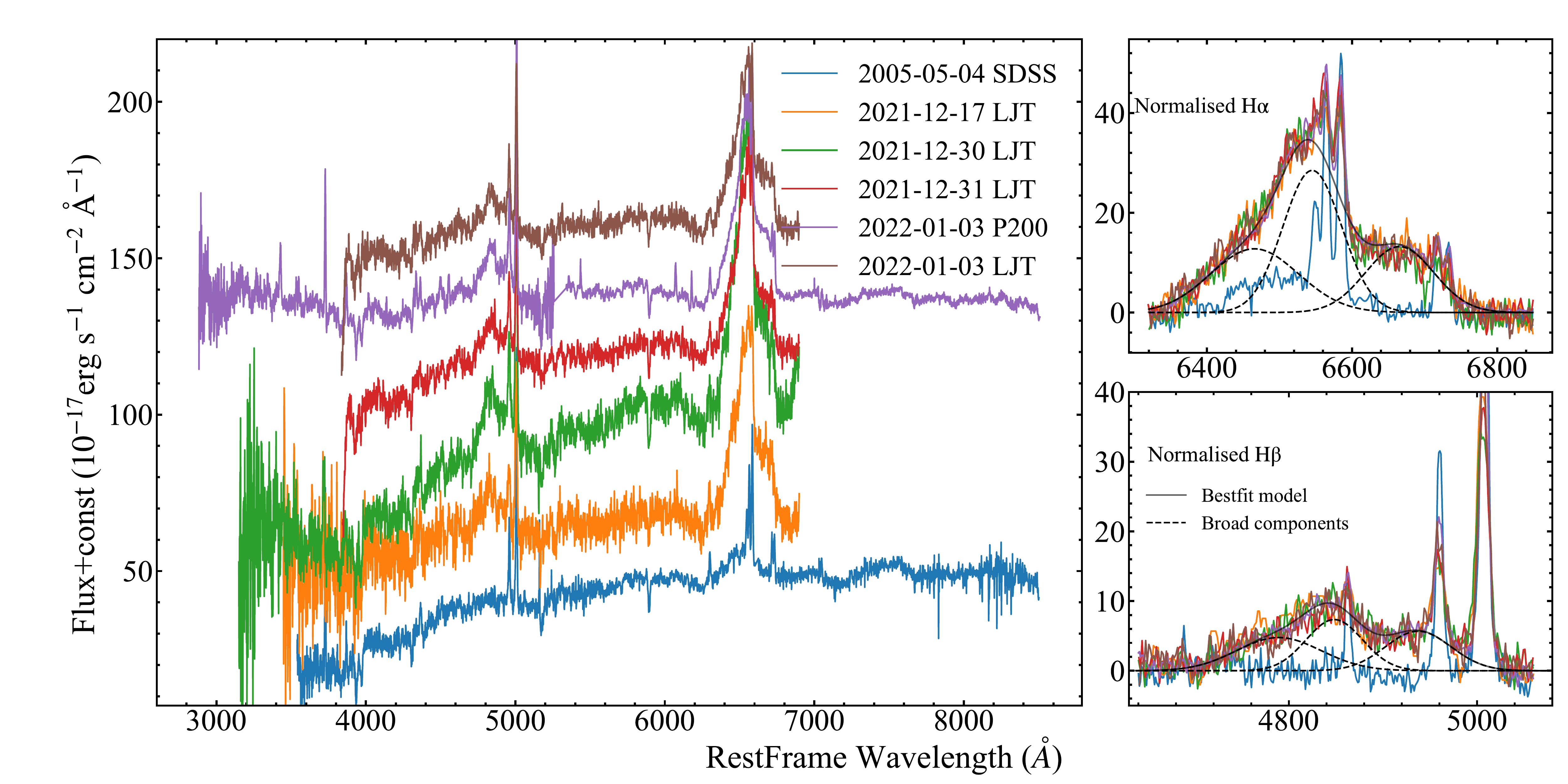}}
\end{minipage}
\caption{
\textbf{The optical follow-up spectra of SDSSJ1430+2303 and their decomposition.}
The spectroscopic observations have been carried out since 2021 Dec. 17 and their reduced data are presented in the left panel. Compared with the SDSS one, they display huge changes in Balmer emissions. The detailed decomposition of the \ha\ and \hb\ emission-line regions are presented in right panels, with color used the same as the left panel. For clarity, we have only overplotted decomposed broad \ha\ and \hb\ components from P200 spectrum taken on 2022 Jan. 3, each of which being fitted by three Gaussian components (black dashed lines).}
\label{1430_spec_follow}
\end{figure*}

\clearpage

\begin{figure*}
\renewcommand{\figurename}{Extended Data Fig.}
\centering
\begin{minipage}{1\textwidth}
\centering{
\includegraphics[angle=0,width=0.36\linewidth]{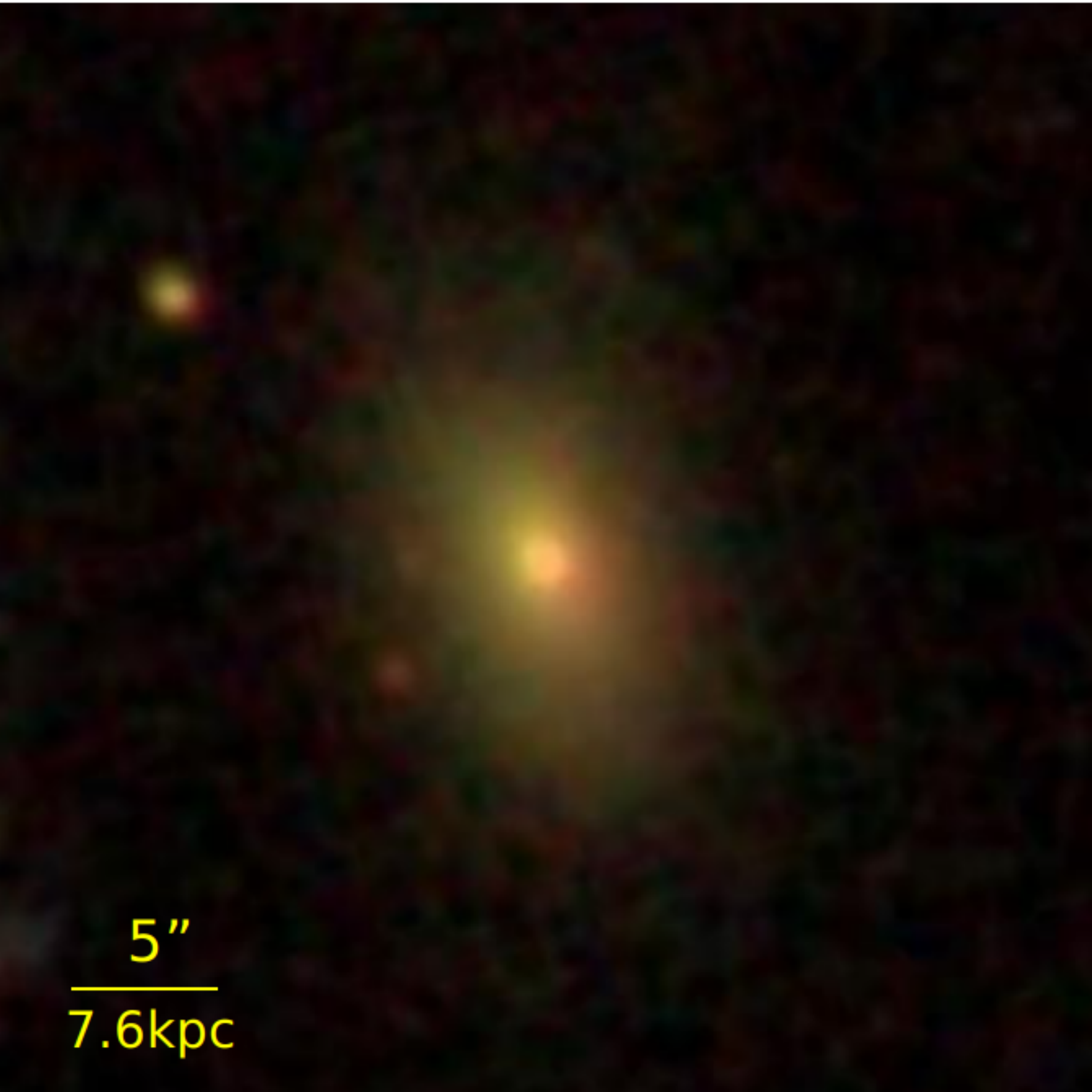}
\includegraphics[angle=0,width=0.62\linewidth]{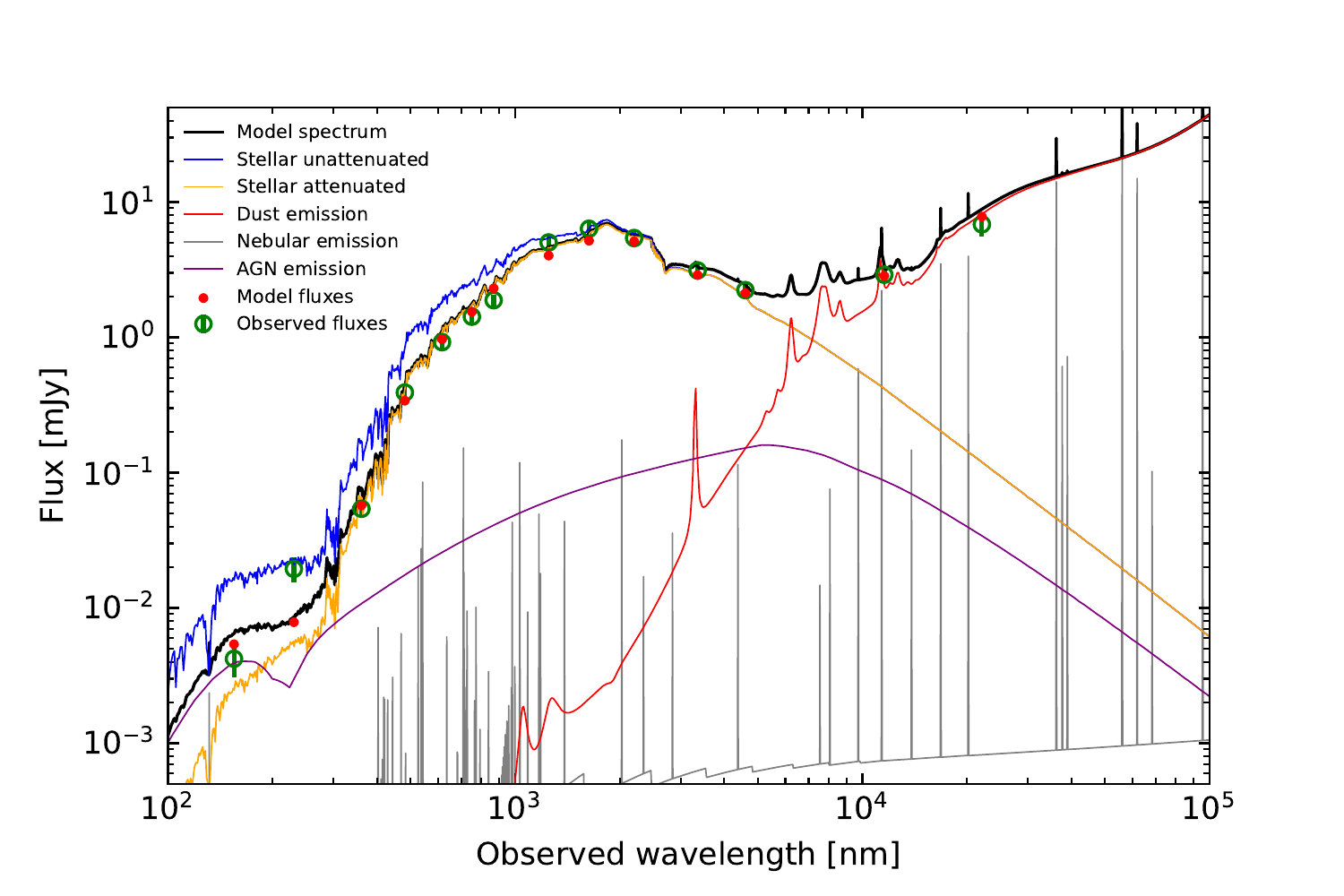}}
\end{minipage}
\caption{
\textbf{Optical image and broad spectral energy distribution (SED) of the host galaxy of SDSSJ1430+2302.} Left: the SDSS $gri$ composited image.
Right: the broadband SED (in flux density per unit frequency in observer's rest frame) fitting with CIGALE. 
}
\label{sed}
\end{figure*}

\clearpage

\begin{figure*}
\renewcommand{\figurename}{Extended Data Fig.}
\centering
\begin{minipage}{1\textwidth}
\centering{\includegraphics[angle=0,width=1.\linewidth]{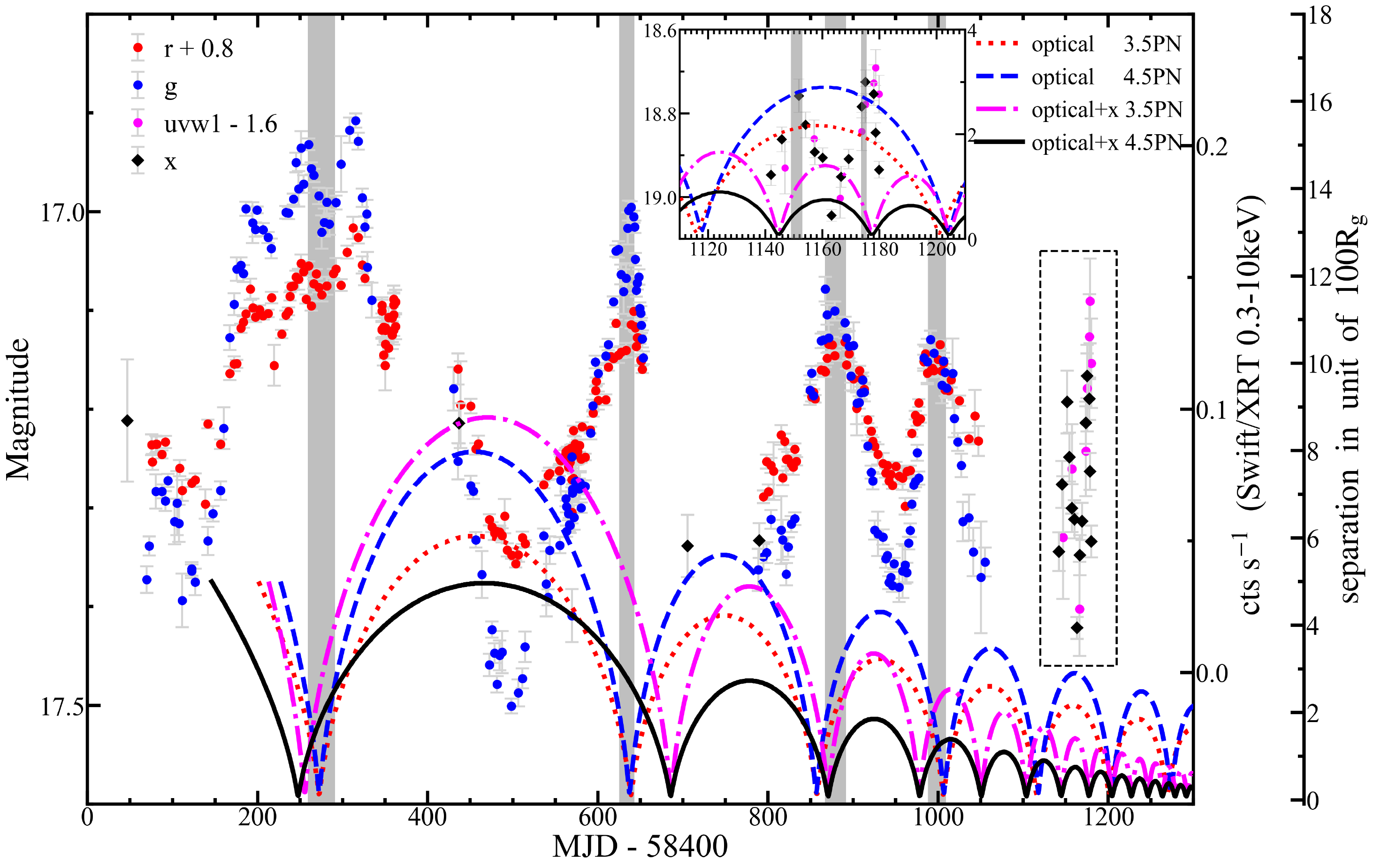}}
\end{minipage}
\caption{\textbf{The evolution of the binary separation compared with the period decaying light curves}. We have overplotted the separations in unit of $100R_g$ of primary SMBH on the light curves as shown in Figure~\ref{lc}, in which the Gaussian fitted peak times, including their uncertainties, are denoted by grey shadow regions. In our trajectory model, we have simply assumed that light curve peaks appear at epochs around pericenter passage. The separations inferred from different scenarios (with or without X-ray peaks, 4.5PN or 3.5PN) are shown in red dotted, blue dashed, magenta dot-dashed and black lines, respectively.
}
\label{lcfit}
\end{figure*}

\clearpage

\begin{figure*}
\renewcommand{\figurename}{Extended Data Fig.}
\centering
\begin{minipage}{1\textwidth}
\centering{\includegraphics[angle=0,width=1.\linewidth]{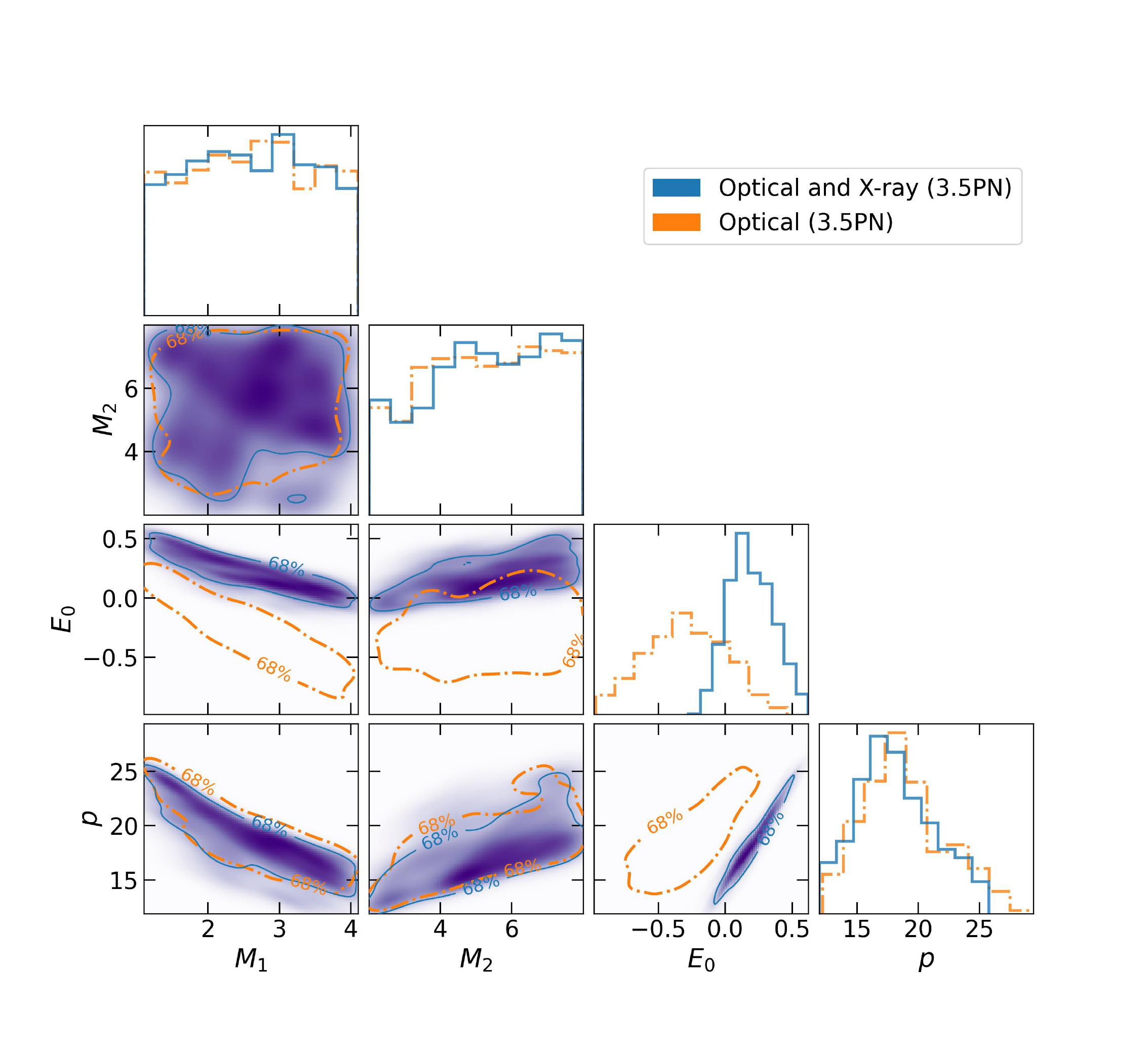}}
\end{minipage}
\caption{\textbf{The posterior distribution of binary parameters, $M_1, M_2, E_0, p$, as inferred from the $3.5$~PN trajectory model}. The Orange contour represents the distribution obtained using only the four optical flare peaks, which shows rather distinct feature from the blue contour, as obtained from both optical and X-ray peaks. In particular, the six-peak Monte-Carlo samples prefer positive energies corresponding to unbound initial orbit.
}
\label{cornerp1}
\end{figure*}

\clearpage

\begin{figure*}
\renewcommand{\figurename}{Extended Data Fig.}
\centering
\begin{minipage}{1\textwidth}
\centering{\includegraphics[angle=0,width=1.\linewidth]{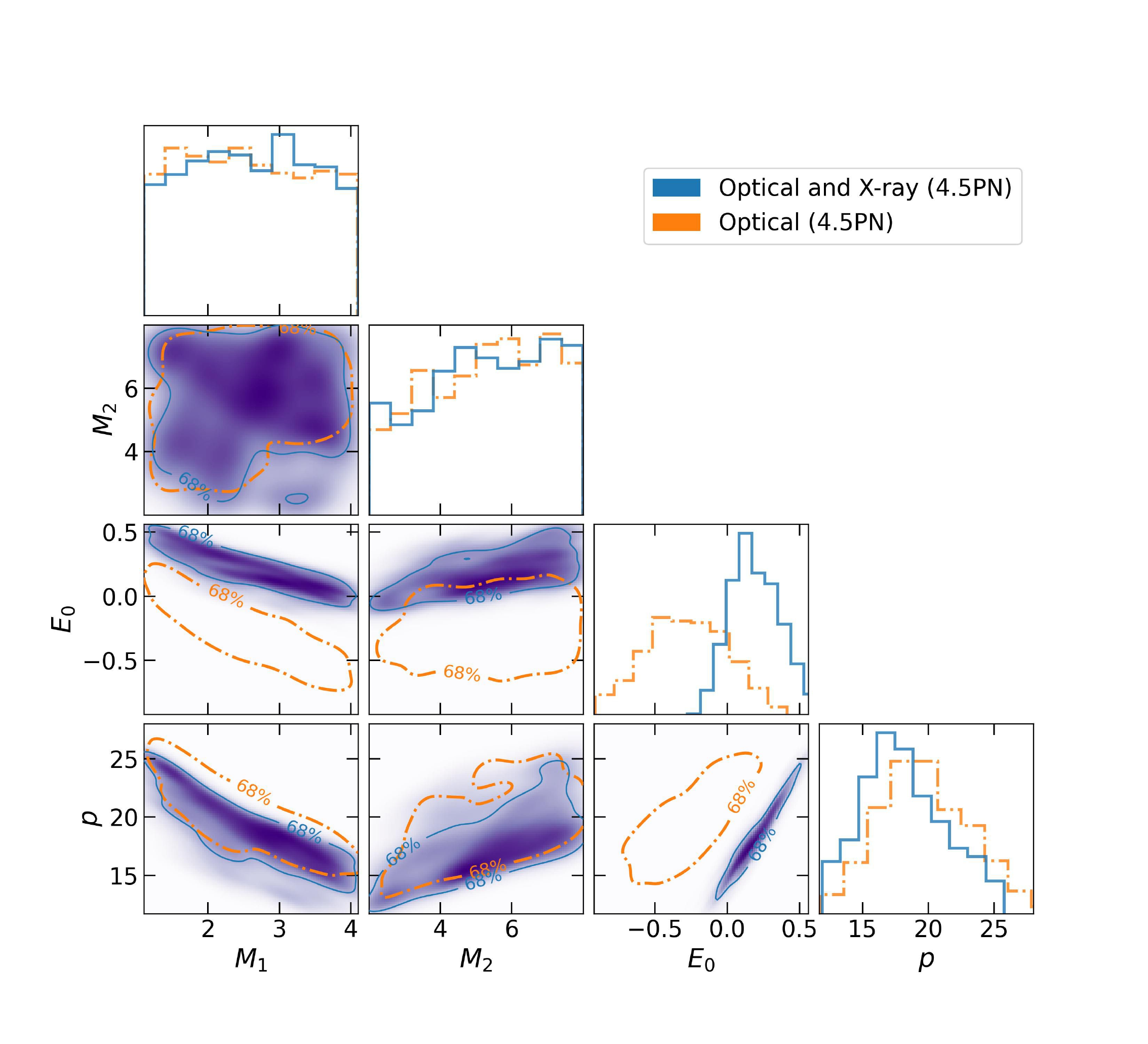}}
\end{minipage}
\caption{\textbf{The posterior distribution of binary parameters, $M_1, M_2, E_0, p$, as inferred from the $4.5$PN trajectory model.} The overall distributions are close to the $3.5$PN case.
}
\label{cornerp2}
\end{figure*}

\clearpage

\begin{figure*}
\renewcommand{\figurename}{Extended Data Fig.}
\centering
\begin{minipage}{1\textwidth}
\centering{\includegraphics[angle=0,width=1.\linewidth]{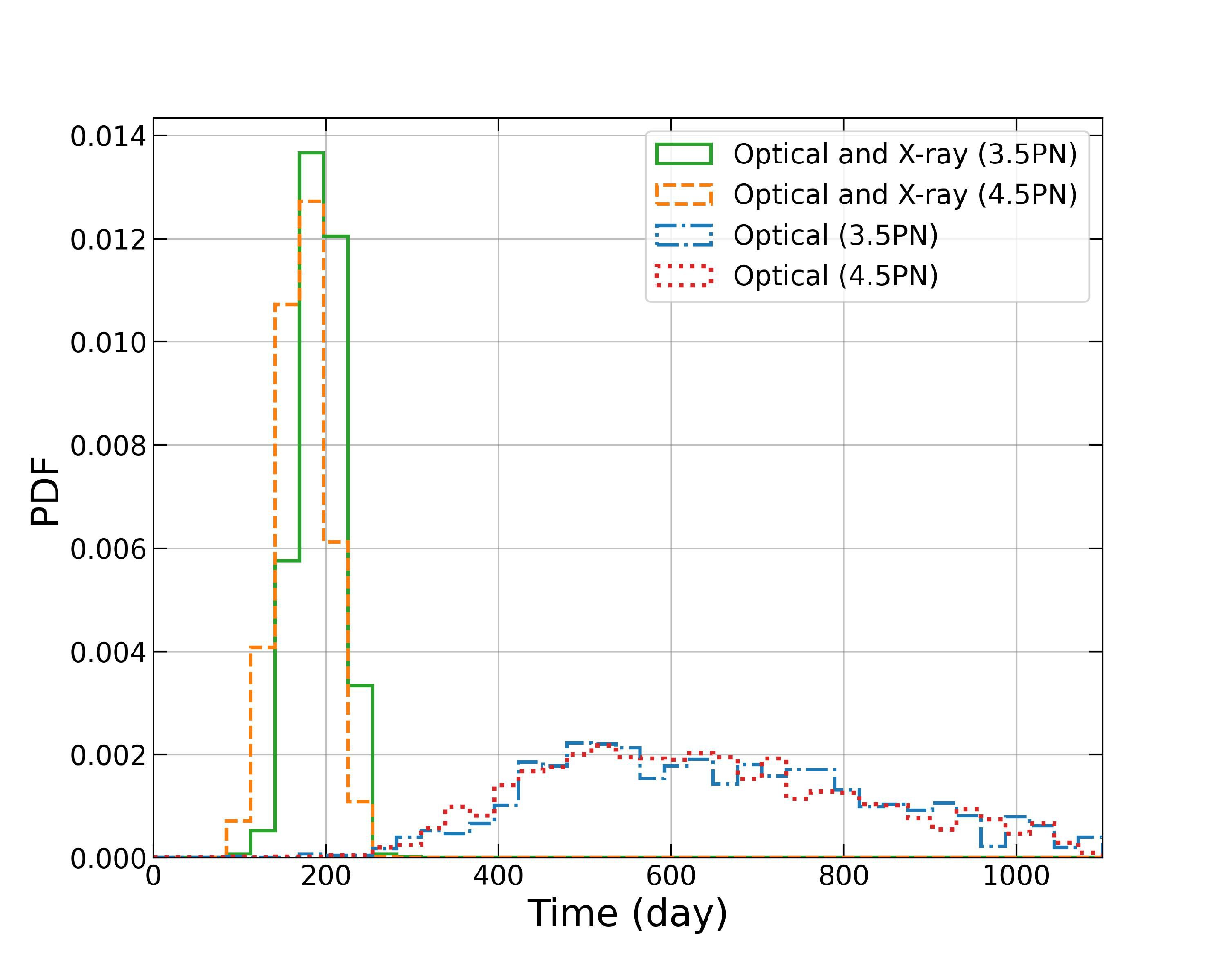}}
\end{minipage}
\caption{\textbf{The posterior distribution of merger time}. The starting point corresponds to the peak time of the second X-ray flare. They are as inferred from the trajectory models, assuming $3.5{\rm PN}/4.5{\rm PN}$ and with/without using the X-ray data. In general the six-peak cases predict imminent merger within one year, and the merger time for four-peak cases is less than three years.
}
\label{mtimepos}
\end{figure*}

\clearpage

\begin{figure}
\renewcommand{\figurename}{Extended Data Fig.}
\centering
\begin{minipage}{1\textwidth}
\centering{\includegraphics[angle=0,width=1.\linewidth]{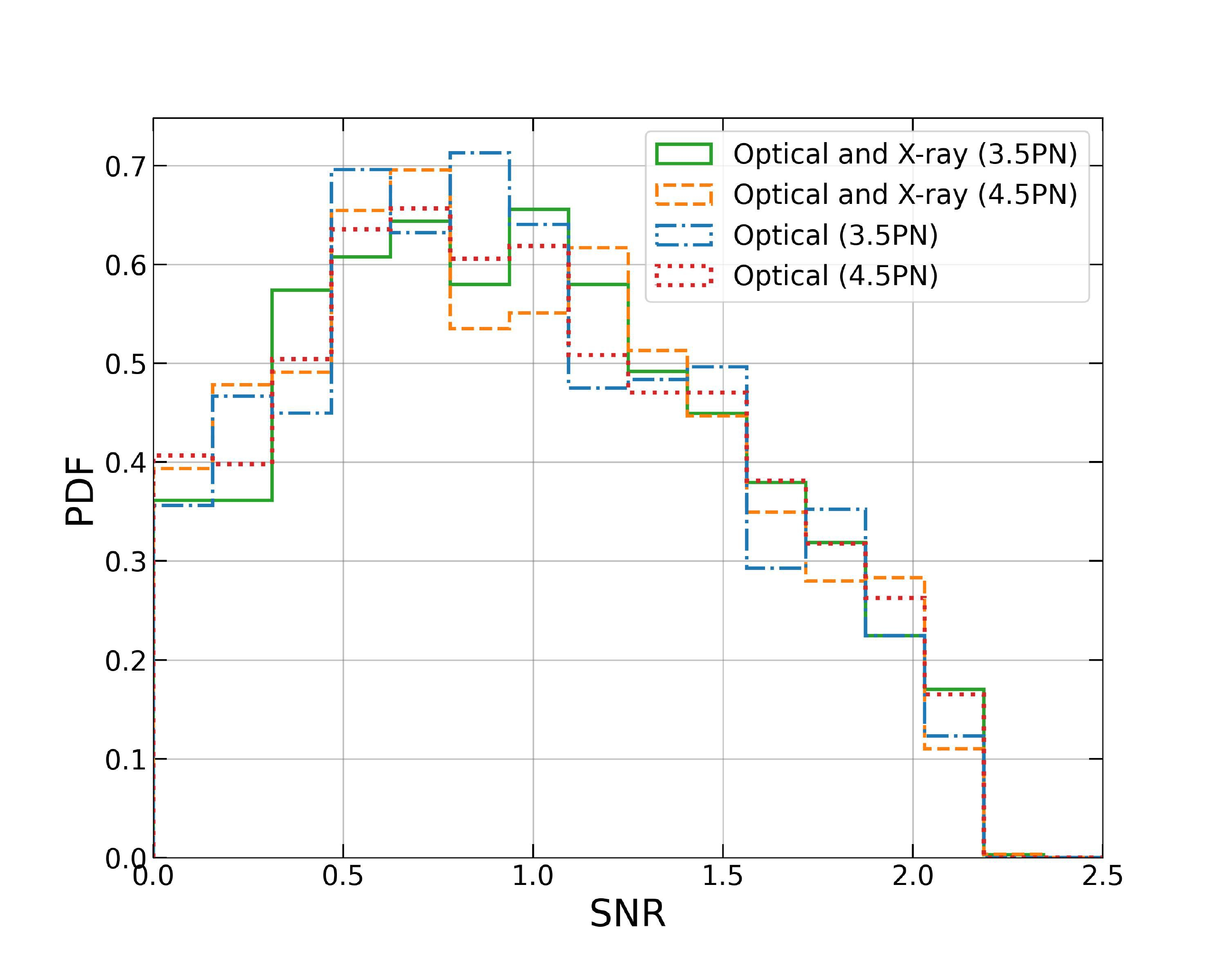}}
\end{minipage}
\caption{\textbf{The expected SNR distrbution for the gravitational memory effect as observed using Pulsar Timing Array, assuming $3.5{\rm PN}/4.5{\rm PN}$ and with/without using the X-ray data.} Without the simplified noise model used in Methods and five year observation, the expected SNR ranges between zero and two. 
}
\label{fig:snr}
\end{figure}

\begin{figure}
\renewcommand{\figurename}{Extended Data Fig.}
\centering
\begin{minipage}{1\textwidth}
\centering{\includegraphics[angle=0,width=1.\linewidth]{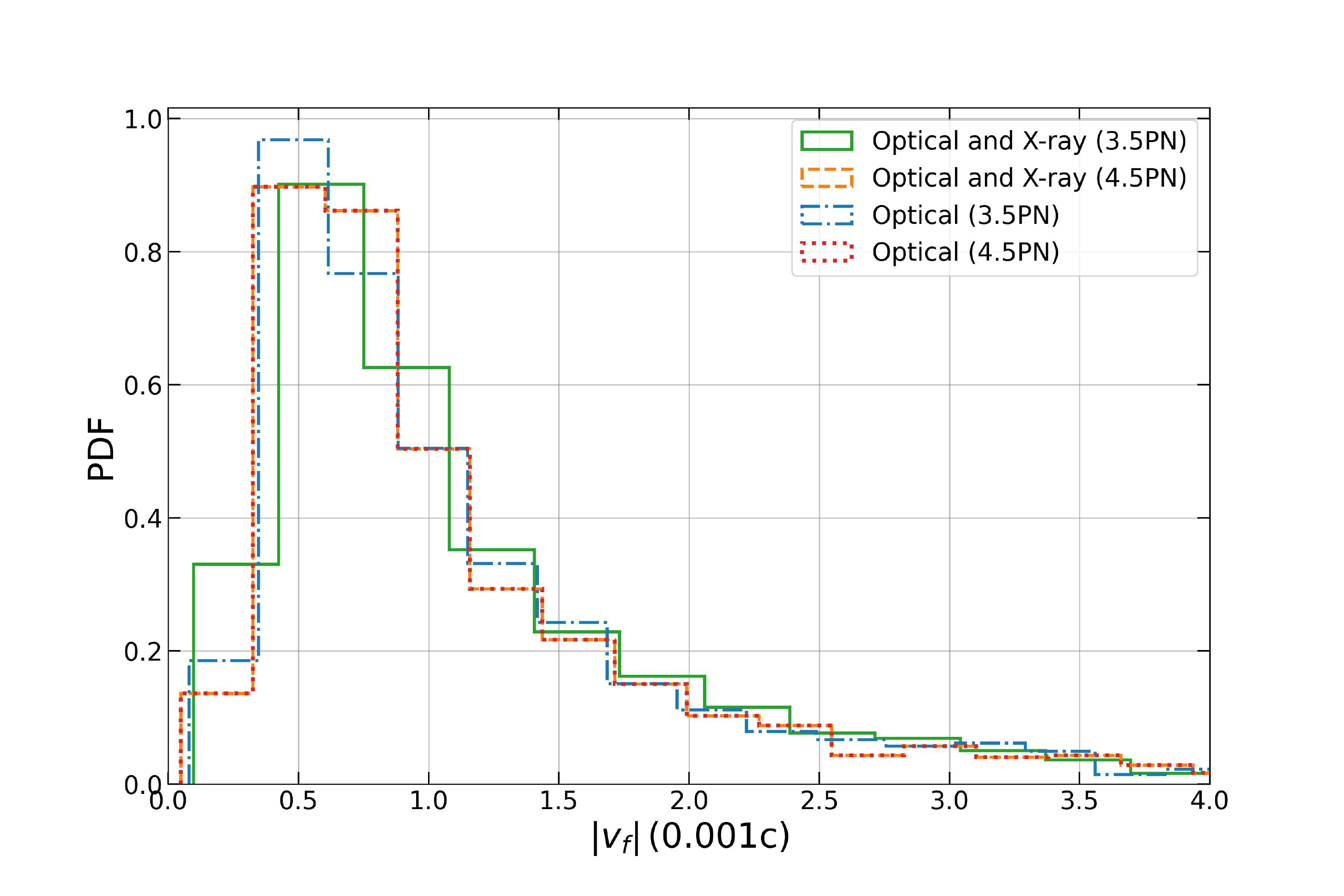}}
\end{minipage}
\caption{\textbf{The expected  distribution of the kick velocity of the final black hole after the binary black hole coalescence}. The dimensionless spin prior for both black holes are assumed to be uniform between zero and one. Because of the relatively large mass ratio, the kick velocity is mostly below $10^3$ ${\rm km s^{-1}}$.
}
\label{fig:kick}
\end{figure}

\clearpage

\begin{table*}[htbp]
\renewcommand{\tablename}{Extended Data Table}
 	\centering  
 	\small
	\caption{Flare Peak Time, luminosity and decay time from Optical and X-ray Measurements} 
	\label{peaktime}
	\begin{tabular}{ccccccc}  
		\hline\hline
	Peak & 1 &  2 &  3 &  4 &  5 &  6   \\
		    \hline
MJD                & 58675.0  & 59031.0 & 59279.5 & 59398.7 & 59552.2 & 59575.2   \\
Uncertainty        & 32.0     & 18.0    & 25.0    & 21.0    & 4.1     & 2.0   \\
$L_g$              & 2.60     & 2.20    & 1.41    & 1.11    & ...     & ...    \\
$t_{g,\rm decay}$  & 20.3     & 16.2    & 20.9    & 21.2    & ...     & ...    \\
$L_r$              & 1.76     & 1.38    & 0.93    & 0.84    & ...     & ...    \\
$t_{r,\rm decay}$  & 23.0     & 17.2    & 26.9    & 19.5    & ...     & ...    \\
$L_x$              & .. & ... & ...  & ...  & 11.66   & 12.4    \\
$t_{x,\rm decay}$  & ...& ... & ...  & ...  & 3.7     & 4.5     \\
        \hline
	\end{tabular}
\begin{tablenotes}
\item 	
{\bf Notes.} The parameters of the first four peaks are extracted from optical light curves and the last two peak are extracted from the X-ray (Swift/XRT) light curve. We have fitted the light curve profiles around each peak with Gaussian functions to get peak time and luminosity (in unit of $10^{43}$~\lum). The decay time is simply defined as the $\sigma$ of Gaussian function fitted to the decaying phases considering their asymmetric profiles. The X-ray luminosity for the latter two peaks are derived from the fitted peak count rate (0.3-10~keV) assuming a spectral shape same to the stacked spectra.
 \end{tablenotes}
\end{table*}

\clearpage

\begin{table*}[htbp]
\renewcommand{\tablename}{Extended Data Table}
\small
 	\centering  
	\caption{Summary of Swift/XRT Results \label{xrt} }  
	\begin{tabular}{ccccc}  
		\hline\hline
ObsID & MJD & Exptime & Count rate & log$L$(0.3-10keV) \\
 &   & second & cts/s & erg/s \\
 \hline
03108725001 & 58446.829 &  177 & 0.0957$\pm$0.0232 & 44.04$\pm$0.26 \\
00012872001 & 58836.955 & 1620 & 0.0946$\pm$0.0076 & 44.04$\pm$0.09 \\
03108725003 & 59105.378 &  344 & 0.0480$\pm$0.0118 & 43.74$\pm$0.27 \\
03108725005 & 59189.635 & 1143 & 0.0501$\pm$0.0066 & 43.76$\pm$0.14 \\
00012872002 & 59542.010 &  851 & 0.0459$\pm$0.0073 & 43.72$\pm$0.17 \\
00012872003 & 59545.860 &  904 & 0.0714$\pm$0.0089 & 43.92$\pm$0.14 \\
00012872005 & 59551.784 &  721 & 0.1027$\pm$0.0119 & 44.07$\pm$0.13 \\
00012872006 & 59554.172 &  948 & 0.0818$\pm$0.0093 & 43.97$\pm$0.12 \\
00012872007 & 59557.415 &  754 & 0.0624$\pm$0.0091 & 43.86$\pm$0.16 \\
00012872008 & 59560.142 & 1121 & 0.0582$\pm$0.0072 & 43.83$\pm$0.13 \\
00012872009 & 59563.255 &  976 & 0.0170$\pm$0.0042 & 43.29$\pm$0.27 \\
00012872010 & 59566.580 &  774 & 0.0446$\pm$0.0076 & 43.71$\pm$0.18 \\
00012872011 & 59569.233 & 1163 & 0.0574$\pm$0.0070 & 43.82$\pm$0.13 \\
00012872013 & 59573.831 & 1468 & 0.0948$\pm$0.0080 & 44.04$\pm$0.09 \\
00012872014 & 59575.159 & 1560 & 0.1127$\pm$0.0085 & 44.11$\pm$0.08 \\
00012872015 & 59577.980 & 1590 & 0.1040$\pm$0.0081 & 44.08$\pm$0.08 \\
00012872016 & 59578.748 & 1553 & 0.0763$\pm$0.0070 & 43.94$\pm$0.10 \\
00012872017 & 59579.913 & 1381 & 0.0497$\pm$0.0060 & 43.76$\pm$0.13 \\
\hline
	\end{tabular}
\end{table*}

\clearpage

\begin{table*}[htbp]
\renewcommand{\tablename}{Extended Data Table}
\small
 	\centering  
	\caption{Details of Optical Spectroscopic Follow-up Observations \label{spec_obs_info} }  
	\begin{tabular}{ccccccc}  
		\hline\hline
Date & Instrument & Grism/Grating &Slit width &Exposure time &wavelength range & S/N$^{b}$           \\
 &     &      &arcsec     &seconds       &\AA            &$\rm pixel^{-1}$\\ \hline
2021-12-17 & YFOSC  & G14 & 2.5 &  2100  &3150-7600  &   22   \\   
2021-12-30 & YFOSC  & G14 & 2.5 &  2100  &3150-7600  &   33   \\   
2021-12-31 & YFOSC  & G14 & 1.8 &  1800  &4300-7600  &   31   \\   
2022-01-03 & DBSP  &600/4000,316/7500 & 1.5 &  2400  &3100-10900 &   102  \\
2022-01-03 & YFOSC  & G14 & 1.8 &  1500  &4300-7600  &   27   \\   
        \hline
	\end{tabular}
    \begin{tablenotes}
    \item 
     $^a$ A UV-blocking filter was added. \\
     $^b$ The S/N was calculated around 6000 \AA\ in restframe.
    \end{tablenotes}
\end{table*}

\end{methods}

\begin{addendum}
\item[Correspondence and request for materials] Ning~Jiang (e-mail: jnac@ustc.edu.cn) and Huan~Yang 

(email:hyang@perimeterinstitute.ca).

\item[Acknowledgements] We thank Yiqiu Ma and Youjun Lu for insightful discussions. We thank Chao-Wei Tsai, Subo Dong, Minghao Yue and Xiaer Zhang for help and efforts on spectroscopic observations. We thank the Swift PI, Brad Cenko and XMM-Newton project scientist, Norbert Schartel for approving the target of opportunity requests.
N.J. acknowledges the financial support by NSFC (11833007, 12073025, 12192221), the B-type Strategic Priority Program of the Chinese Academy of Sciences (Grant No. XDB41000000), and China Manned Spaced Project (CMS-CSST-2021-B11). H.Y. and Z.P. acknowledge supports by the Natural Sciences and Engineering Research Council of Canada and in part by Perimeter Institute for Theoretical Physics. Research at Perimeter Institute is supported in part by the Government of Canada through the Department of Innovation, Science and Economic Development Canada and by the Province of Ontario through the Ministry of Colleges and Universities. Z.Z. acknowledges the support by NSFC (12022303). This work is based on observations made with ZTF, Swift, XMM-Newton, LJT and P200. This research uses data obtained through the Telescope Access Program (TAP). Observations obtained with the Hale Telescope at Palomar Observatory were obtained as part of an agreement between the National Astronomical Observatories, Chinese Academy of Sciences, and the California Institute of Technology. The ZTF forced-photometry service was funded under the Heising-Simons Foundation grant \#12540303 (PI: Graham).

\item[Author contributions] 
N.J. led the observational discovery and follow-ups.
H.Y. led the theoretical interpretation of the discovery.
J.Z. analyzed the optical data and Swift/UVOT data.
Z.L. performed the trajectory model simulation and fitting.
L.D. analyzed the Swift/XRT and XMM-Newton data.
Y.W. analyzed the optical spectroscopic data.
J.W. arranged and performed the LJT observations.
Z.P. performed the flare model calculations.
H.L. performed the host SED fitting.
L.D., X.S. and T.W. contributed to propose the Swift and XMM-Newton observations.
N.J. and H.Y. jointly drafted the manuscript.
T.W. and Z.Z. provided valuable feedback and all co-authors helped to shape the manuscript.

\item[Competing Interests] The authors declare that they have no
competing financial interests.
\end{addendum}

\bibliography{reference}

\end{document}